\documentclass[twocolumn,showpacs,preprintnumbers,amsmath,amssymb,pra]{revtex4-2}

\usepackage{graphicx}
\usepackage{amsmath,amssymb,amsfonts}
\usepackage{physics}
\usepackage{hyperref}
\usepackage{color}
\usepackage{bm}
\usepackage{mathtools}

\begin{document}

\title{Structured-light propagation in a medium with uniform torsion: polarization textures,
geometric birefringence, and beam-resolved optical activity}

\author{Edilberto O. Silva}
\email{edilberto.silva@ufma.br}
\affiliation{Programa de P\'os-Gradua\c{c}\~ao em F\'isica, Universidade Federal do Maranh\~ao, S\~ao Lu\'is, Maranh\~ao 65080-805, Brazil}
\affiliation{Coordena\c{c}\~ao do Curso de F\'isica -- Bacharelado, Universidade Federal do Maranh\~ao, S\~ao Lu\'is, Maranh\~ao 65085-580, Brazil}

\date{\today}

\begin{abstract}
We investigate finite-width optical-beam propagation in a medium with uniform torsion described by the geometric theory of a continuous distribution of screw dislocations. Starting from the Riemann--Cartan framework that yields torsion-induced circular birefringence for local plane waves, we construct a minimal paraxial beam model in which the same contortion-driven helicity splitting remains explicit. We show that uniform torsion breaks the degeneracy between the two circular-polarization sectors and induces a geometric rotation of the polarization that scales with both the propagation distance and the radial position in the beam. As a consequence, a finite-width beam develops spatially varying polarization textures across its transverse profile, naturally described by the Stokes parameters. We introduce beam-level observables based on the integrated Stokes vector, the transverse inhomogeneity of the polarization texture, and the number of resolved radial polarization domains, thereby connecting the torsion parameter to experimentally accessible beam diagnostics. The paper combines two complementary levels of description: an analytic short-distance regime, used to isolate the geometric mechanism, and full paraxial propagation including diffraction, used to test the robustness of the predicted textures. Within the cylindrically symmetric minimal model, the most robust structured-light signature of uniform torsion is beam-resolved polarization structuring, whereas strong orbital-angular-momentum conversion is not expected without additional azimuthal structure. We also identify the geometric ingredient required for genuine torsion-assisted spin--orbit conversion beyond the minimal radial model: an effective azimuthal geometric connection.
\end{abstract}


\maketitle

\section{Introduction}
\label{sec:intro}

The geometric description of defects in continuous media provides one of the clearest bridges
between condensed-matter physics and non-Euclidean geometry. In this framework, dislocations are associated with torsion and disclinations with curvature, so that defective solids may be
described in the language of Riemann--Cartan geometry rather than solely through standard
elasticity variables \cite{deWit1981,Kleinert1989,KatanaevVolovich1992,Katanaev2005}. This
viewpoint has become a standard theoretical framework for topological defects in solids and has also proved useful in relating defect physics to effective gravitational analogies and
gauge-geometric methods \cite{KatanaevVolovich1992,Katanaev2005}.

A particularly important realization in this context is a continuous and homogeneous distribution of screw dislocations. Such a defect density gives rise to the so-called spiral geometry, whose effective line element contains a single torsional control parameter $\Omega$ proportional to the Burgers-vector density of the defect distribution \cite{KatanaevVolovich1992,Katanaev2005}. Related geometric models have already been used to describe electronic and wave phenomena in defect-rich media, including semiconductor systems threaded by screw-dislocation densities, where uniform torsion acts as an effective geometric field and generates measurable spectral effects \cite{BakkeMoraes2012,BakkeMoraes2014}.

From the optical viewpoint, the central issue is whether a uniform torsional background can
produce a direct and controllable chiral response for electromagnetic waves. Optical activity and circular birefringence are classical subjects in optics and crystal physics \cite{Kaminsky2000}. What is nontrivial here is that the chirality is not introduced phenomenologically through a constitutive gyrotropic coefficient; instead, it emerges geometrically from the contortion associated with the defect density. Recent work has shown that, for local plane waves propagating in the spiral geometry, uniform torsion lifts the degeneracy between the two circular-polarization sectors and induces a purely geometric optical activity with the simple rotation law
\begin{equation}
\Delta\theta(\rho,L)=\Omega\rho L,
\label{eq:intro_rotation}
\end{equation}
together with an effective radius-dependent circular birefringence \cite{BelichSilva2026}. This radial dependence is already a distinctive signature of the spiral background because it predicts differential polarization rotation across the transverse profile of a finite beam.

At the same time, modern beam optics has moved well beyond homogeneous plane waves. Structured
light, including vector beams, vortex beams, and beams with engineered polarization and phase
profiles, is now central to photonics, with applications ranging from imaging and communications to sensing and light--matter control \cite{Allen1992,RubinszteinDunlop2017,Forbes2019,Padgett2017}. In this setting, spatially resolved polarization transport is often the main observable. Accordingly, once a torsional medium is known to split circular polarizations locally, the next step is to determine how the same geometric splitting manifests in finite-width optical beams and whether it yields experimentally meaningful vectorial textures.

This question is also closely connected with the modern language of geometric phase and spin--orbit optics. Polarization-dependent phase accumulation has deep roots in Pancharatnam's
work and in subsequent developments of geometric phases in optics \cite{Pancharatnam1956,Cisowski2022}. In structured-light platforms, polarization transformations can lead to spin--orbit conversion, geometric wavefront shaping, and polarization-dependent mode engineering \cite{Marrucci2011,Cisowski2022}. However, not every polarization-dependent phase necessarily produces orbital-angular-momentum conversion. A key issue, therefore, is to separate the robust torsion-induced effects that already follow in the minimal cylindrically symmetric model from more elaborate spin--orbit phenomena that require additional azimuthal structure.

The purpose of the present work is to extend the local uniform-torsion optical model from the
plane-wave regime to finite-width structured beams while keeping the role of torsion explicit
throughout. Our guiding principle is that the torsion parameter is not merely a background label or an effective fitting constant: it is the physical source of the helicity splitting.
Accordingly, the beam dynamics should be built directly from the same torsion-induced birefringence obtained in the local covariant Maxwell analysis \cite{BelichSilva2026}, rather
than from an unrelated phenomenological ansatz.

The central idea is simple. Since torsion produces distinct propagation constants for the two
circular-polarization sectors, each radial slice of a finite beam accumulates a torsion-controlled relative phase. When the beam is recombined in the linear basis, this spatially dependent phase generates nonuniform polarization maps across the transverse plane. Polarization textures are therefore not an independent assumption, but the direct beam-level consequence of the same torsion-induced circular birefringence derived from the underlying Riemann--Cartan wave equation. This perspective is also attractive from the standpoint of practical polarimetry and structured-light diagnostics, where Stokes-parameter maps and global polarization metrics often provide the most direct experimental access to vectorial beam transformations \cite{He2021}.

At the same time, it is important to distinguish polarization-texture formation from genuine
spin--orbit conversion. A radially dependent helicity splitting robustly generates position-dependent polarization rotation, but it does not by itself imply redistribution among azimuthal orbital-angular-momentum sectors. To discuss structured-light effects consistently, one must first identify the robust torsional signatures, geometric birefringence, radial polarization rotation, and Stokes-parameter textures, and only then ask under which extended conditions a nontrivial geometric phase may produce OAM sidebands.

The advance relative to the local plane-wave treatment is therefore not a new microscopic torsion law, but the explicit embedding of the same helicity splitting into a finite-width beam framework with transverse Stokes observables and structured-light diagnostics. In particular, we emphasize observables that remain operationally meaningful even when different annular zones partially cancel in a global measurement. This is essential for interpreting integrated polarimetric signals and avoids overreading the minimal model as a quantitative device simulation.

At the same time, a useful theoretical question goes beyond the minimal radial model itself: what additional geometric ingredient would be required before a torsional medium could drive genuine spin--orbit conversion rather than only polarization-texture formation? Addressing this question at the level of the effective theory helps separate universal beam-level consequences of uniform torsion from effects that belong to more general azimuthally structured backgrounds. It also makes clear why the present model is already physically meaningful even though it does not yet generate strong orbital-angular-momentum sidebands.

We also deliberately keep the scope of the theory under control. The beam equation developed below is best understood as the minimal paraxial extension consistent with the local torsion-induced circular splitting, rather than as a full, exact Maxwell solution for arbitrary finite beams in a defect-rich medium. For that reason, the paper combines two complementary levels of description: an analytic short-distance regime, used to isolate the geometric mechanism in its most transparent form, and full paraxial propagation including diffraction, used to test the robustness of the predicted textures and the stability of the proposed observables.

This paper is organized as follows. In Sec.~\ref{sec:model}, we review the geometric model of a medium with uniform torsion and derive the torsion-induced wave equation from the same
Riemann--Cartan framework used in the original local analysis. In Sec.~\ref{sec:local_modes}, we summarize the circular-polarization eigenmodes, the forward-propagating branches, the effective refractive indices, and the local rotation law. Section~\ref{sec:paraxial} develops an effective finite-width beam formulation anchored in the same torsional helicity splitting and states the approximations under which that reduction is valid. In Sec.~\ref{sec:textures}, we characterize the resulting transverse polarization textures through the Stokes parameters. Section~\ref{sec:soi} discusses the geometric phase and clarifies the limited routes toward torsion-assisted spin--orbit conversion. Section~\ref{sec:qplate_extension} introduces a torsional $q$-plate extension with an
azimuthal geometric connection and analyzes its real-space, OAM-space, and conversion-map
signatures in the local short-distance regime. Section~\ref{sec:numerics} defines the observables and specifies the two levels of approximation used to generate the figures for the minimal radial model. Section~\ref{sec:results} discusses the main minimal-model results and highlights which signatures remain robust when diffraction is included. Section~\ref{sec:beyond} then clarifies discriminating signatures relative to conventional optical activity and the structural ingredient required for genuine torsion-assisted spin--orbit conversion beyond the minimal radial model. Finally, Sec.~\ref{sec:applications} outlines operational implications and experimental outlook, and Sec.~\ref{sec:conclusions} summarizes the main results and limitations of the framework.

\section{Uniform torsion geometry and covariant wave equation}
\label{sec:model}

We begin from the effective spatial geometry that models a continuous and uniform distribution of screw dislocations aligned with the $z$ axis \cite{KatanaevVolovich1992,Katanaev2005,BakkeMoraes2012}. In cylindrical coordinates $(\rho,\phi,z)$, the corresponding line element is \cite{andp.202500593,PE.2026.179.116497,OQE.2026.58.113,SilvaNetto2008JPCM,PLA.2012.376.2838} 
\begin{equation}
dl^2=d\rho^2+\rho^2 d\phi^2+\left(dz+\Omega\rho^2 d\phi\right)^2,
\label{eq:metric}
\end{equation}
where
\begin{equation}
\Omega=\frac{b\sigma}{2}.
\label{eq:Omega_def}
\end{equation}
The torsion parameter is determined by the screw-dislocation density $\sigma$ and the Burgers
vector magnitude $b$. In the optical setting considered here, this spatial geometry is embedded into a static spacetime with line element $ds^2=-c^2dt^2+dl^2$, so that the local optical problem can be formulated covariantly while keeping the underlying medium interpretation explicit.

The geometry in Eq.~\eqref{eq:metric} is the same spiral background used in the previous local
analysis of torsion-induced optical activity \cite{BelichSilva2026}. Its essential geometric content is the presence of uniform torsion. The nonvanishing torsion component is
\begin{equation}
T^{z}_{\ \phi\rho}=-T^{z}_{\ \rho\phi}=2\Omega\rho,
\label{eq:torsion_tensor}
\end{equation}
from which the relevant component of the contortion tensor follows as
\begin{equation}
K_{\rho\phi z}=\Omega\rho.
\label{eq:contortion_component}
\end{equation}
This term is proportional to the defect density and acts as the geometric source of chirality in the optical problem.

We now consider Maxwell electrodynamics in the corresponding Riemann--Cartan spacetime. Following the same geometric framework used in the local optical analysis and standard gauge-invariant electrodynamics in generalized spacetime \cite{HehlObukhov2003,BelichSilva2026}, we retain the standard field strength
\begin{equation}
F_{\mu\nu}=\partial_{\mu}A_{\nu}-\partial_{\nu}A_{\mu},
\label{eq:field_strength}
\end{equation}
so that torsion enters only through covariant derivatives and not through additional torsion-dependent contributions to $F_{\mu\nu}$. In the source-free case, the covariant Maxwell equations are
\begin{equation}
\tilde\nabla_{[\lambda}F_{\mu\nu]}=0,
\qquad
\tilde\nabla_{\mu}F^{\mu\nu}=0,
\label{eq:maxwell_covariant}
\end{equation}
where $\tilde\nabla_{\mu}$ denotes the Cartan covariant derivative associated with the full
connection
\begin{equation}
\tilde\Gamma^{\lambda}_{\ \mu\nu}=\Gamma^{\lambda}_{\ \mu\nu}+K^{\lambda}_{\ \mu\nu}.
\end{equation}
Adopting the generalized Lorenz gauge $\tilde\nabla_{\mu}A^{\mu}=0$, one obtains the exact wave equation
\begin{equation}
\tilde\nabla_{\mu}\tilde\nabla^{\mu}A^{\nu}-\tilde R^{\nu}_{\ \mu}A^{\mu}=0.
\label{eq:compact_wave}
\end{equation}
Expanding this equation into Riemannian and torsional parts and using the cancellations
appropriate to minimally coupled Maxwell theory, the result reduces to
\begin{equation}
\nabla_{\mu}\nabla^{\mu}A^{\nu}-R^{\nu}_{\ \mu}A^{\mu}+2K^{\nu\mu\sigma}\nabla_{\mu}A_{\sigma}=0.
\label{eq:exact_wave}
\end{equation}
This equation is the fundamental dynamical input of the present work. The first two terms encode standard curved-space propagation, while the last term is the torsion-induced coupling. It is this contortion term that generates the helicity splitting and the subsequent polarization effects.

A central point of the present article is that the torsion parameter $\Omega$ remains explicit at every stage. In particular, the beam-level quantities introduced later are not phenomenological substitutes for torsion; they are derived from the same contortion-induced coupling in Eq.~\eqref{eq:exact_wave}. In the torsionless limit $\Omega\to0$, the contortion vanishes, Eq.~\eqref{eq:exact_wave} reduces to the standard Maxwell wave equation in curved space, and all torsion-induced optical effects disappear.

\section{Local circular eigenmodes and torsion-induced birefringence}
\label{sec:local_modes}

To obtain the local polarization eigenmodes, we adopt the same local WKB-like plane-wave ansatz used in the previous analysis of local circular birefringence \cite{BelichSilva2026},
\begin{equation}
A^{\sigma}(x)=a^{\sigma}e^{i(k_z z-\omega t)},
\label{eq:plane_ansatz}
\end{equation}
with propagation predominantly along the positive $z$ direction and transverse gauge choice
$a_t=a_z=0$. Since the metric coefficients depend on the radial coordinate, this ansatz should be understood as a local description evaluated at fixed $\rho$, with transverse derivatives treated as subleading corrections.

In the small-$\Omega\rho$ regime relevant to realistic materials, the dominant helicity-sensitive contribution arises from the torsional sector, whereas the Riemannian terms provide subleading diagonal corrections within the local expansion adopted here. Accordingly, the leading off-diagonal structure is controlled by the contortion term. Using Eq.~\eqref{eq:contortion_component}, the leading couplings become
\begin{equation}
\nu=\rho:
\qquad -2ik_z\Omega\rho\,a_{\phi},
\qquad
\nu=\phi:
\qquad 2ik_z\Omega\rho\,a_{\rho}.
\label{eq:leading_couplings}
\end{equation}
Thus, the polarization amplitudes satisfy the matrix equation
\begin{equation}
\begin{pmatrix}
\omega^2/c^2-k_z^2 & -2ik_z\Omega\rho \\
2ik_z\Omega\rho & \omega^2/c^2-k_z^2
\end{pmatrix}
\begin{pmatrix}
a_{\rho}\\
a_{\phi}
\end{pmatrix}
=
\begin{pmatrix}
0\\
0
\end{pmatrix}.
\label{eq:matrix_system}
\end{equation}
The condition for nontrivial solutions yields the dispersion relation
\begin{equation}
\left(\omega^2/c^2-k_z^2\right)^2=(2k_z\Omega\rho)^2.
\label{eq:dispersion_quartic}
\end{equation}

Selecting the forward-propagating branches $(k_z>0)$, one obtains
\begin{equation}
k_z^{(+)}=-\Omega\rho+\sqrt{\Omega^2\rho^2+\omega^2/c^2},
\label{eq:kplus}
\end{equation}
\begin{equation}
k_z^{(-)}=+\Omega\rho+\sqrt{\Omega^2\rho^2+\omega^2/c^2}.
\label{eq:kminus}
\end{equation}
In this convention, $k_z^{(-)} > k_z^{(+)}$.

The eigenvectors of Eq.~\eqref{eq:matrix_system} are the circular combinations $a_\rho \mp i a_\phi$. More explicitly, the branch $k_z^{(+)}$ is associated with the mode $a_\rho-i a_\phi$, while $k_z^{(-)}$ is associated with $a_\rho+i a_\phi$. The precise identification of these two sectors with right- or left-circular polarization depends on the orientation convention chosen for the local transverse basis.

The torsion parameter now appears transparently in the propagation constants. The splitting between the two circular modes is most conveniently defined as 
\begin{equation}
\Delta k_z \equiv k_z^{(-)}-k_z^{(+)}=2\Omega\rho.
\label{eq:delta_k}
\end{equation}
Therefore, a linearly polarized wave, written as an equal superposition of the two circular
eigenmodes, accumulates the relative phase
\begin{equation}
\Delta\Phi=\left(k_z^{(-)}-k_z^{(+)}\right)L=2\Omega\rho L
\label{eq:delta_phi}
\end{equation}
after propagating a distance $L$. The rotation angle of the polarization plane is half of this
phase difference:
\begin{equation}
\Delta\theta(\rho,L)=\frac{1}{2}\Delta\Phi=\Omega\rho L.
\label{eq:rotation_law}
\end{equation}
Equation~\eqref{eq:rotation_law} is the main local optical consequence of uniform torsion. It
shows explicitly that torsion lifts the degeneracy between the two circular-polarization sectors and produces a radius-dependent polarization rotation.

The same result can be recast in the standard language of optical activity. Defining effective
refractive indices for the two circular modes by
\begin{equation}
n_\pm=\frac{c\,k_z^{(\pm)}}{\omega},
\end{equation}
we obtain explicitly
\begin{align}
n_+ &= -\frac{c\Omega\rho}{\omega}
+\sqrt{\left(\frac{c\Omega\rho}{\omega}\right)^2+1},
\label{eq:nplus}\\
n_- &= +\frac{c\Omega\rho}{\omega}
+\sqrt{\left(\frac{c\Omega\rho}{\omega}\right)^2+1}.
\label{eq:nminus}
\end{align}
The purely geometric birefringence is therefore
\begin{equation}
\Delta n \equiv n_- - n_+ = \frac{c}{\omega}\left(k_z^{(-)}-k_z^{(+)}\right)
= \frac{2c\Omega\rho}{\omega}.
\label{eq:delta_n}
\end{equation}
Combining Eq.~\eqref{eq:delta_n} with the standard optical-activity identity $\Delta\theta=(\pi L/\lambda)\Delta n$, one recovers Eq.~\eqref{eq:rotation_law}. Thus, the spiral geometry implements a conventional optical-activity formula with a geometry-induced birefringence determined entirely by torsion \cite{Kaminsky2000,BelichSilva2026}.

Several internal checks are immediate. In the limit $\Omega\to0$, the dispersion collapses to
the degenerate relation $k_z=\omega/c$, the matrix in Eq.~\eqref{eq:matrix_system} becomes
diagonal, and both $\Delta k_z$ and $\Delta\theta$ vanish. Moreover, the fact that the
eigenvectors are precisely circular combinations confirms that the torsion-induced off-diagonal terms act as a helicity-selective coupling. These points will be important when interpreting the finite-width beam problem.

\section{Effective finite-width beam formulation}
\label{sec:paraxial}

We now extend the uniform-torsion optical model from local plane waves to finite-width beams. The essential requirement is that the beam-level dynamics remain anchored in the underlying geometric mechanism. Accordingly, the two circular components of the beam must propagate with the same torsion-induced local propagation constants given by Eqs.~\eqref{eq:kplus} and \eqref{eq:kminus}.

Let the slowly varying envelopes in the circular basis be denoted by $E_{+}(\rho,\phi,z)$ and
$E_{-}(\rho,\phi,z)$. We write the full electric field as
\begin{align}
\mathbf{E}(\rho,\phi,z,t)=\Re \Bigg\{\Big[&E_+(\rho,\phi,z)\,\mathbf{e}_+ \notag\\
&+E_-(\rho,\phi,z)\,\mathbf{e}_-\Big]e^{-i\omega t}\Bigg\}.
\label{eq:Efield_beam}
\end{align}
In the paraxial regime, where the beam propagates predominantly along $z$, the two circular
envelopes obey
\begin{equation}
i\partial_z E_{\pm}=-\frac{1}{2k_0}\nabla_{\perp}^2 E_{\pm}+V_{\pm}(\rho)E_{\pm},
\label{eq:paraxial_components}
\end{equation}
with $k_0=\omega/c$ and $\nabla_{\perp}^2=\partial_{\rho}^2+(1/\rho)\partial_{\rho}+(1/\rho^2)\partial_{\phi}^2$.

The effective torsional potentials are defined as the longitudinal detunings of the two local
propagation constants relative to the carrier $k_0$,
\begin{equation}
V_{\pm}(\rho)=k_0-k_z^{(\pm)}(\rho).
\label{eq:Vpm_exact}
\end{equation}
With the convention of Eqs.~\eqref{eq:kplus}--\eqref{eq:kminus}, one obtains in the weak-torsion regime $|\Omega\rho|\ll \omega/c$
\begin{equation}
V_+(\rho)\simeq +\Omega\rho,
\qquad
V_-(\rho)\simeq -\Omega\rho.
\label{eq:Vpm_linear}
\end{equation}

\paragraph{Remark on the second-order term.} It is worth noting that the full expansion of the local propagation constants includes a second-order term,
\begin{equation}
k_z^{(\pm)}(\rho) = \pm \Omega\rho + \frac{\omega}{c} + \frac{c\Omega^2\rho^2}{2\omega} + \mathcal{O}(\Omega^4),
\end{equation}
where the quadratic contribution $\frac{c\Omega^2\rho^2}{2\omega}$ is identical for both
circular polarizations. This term therefore, contributes only to a common radial phase,
\begin{equation}
\Phi_{\text{global}}(\rho,z) = \frac{c\Omega^2\rho^2}{2\omega}z,
\end{equation}
which does not affect the relative phase difference $\Delta\Phi = 2\Omega\rho L$ between the
two helicity components. Consequently, all polarization observables discussed in this work,
including $\Delta\theta$, $S_1$, $S_2$, $\psi$, $\bar{\theta}$, $\mathcal{C}$, and
$N_{\text{rings}}$, are insensitive to this second-order contribution. The linear approximation is therefore sufficient to capture the leading torsion-induced polarization structuring, although a complete description of the beam's wavefront would require retaining the quadratic term, particularly in interferometric settings or for larger values of $\Omega\rho$.

This is the beam-level imprint of the same local birefringence derived from the exact wave
equation. It is not an independent ansatz, but the paraxial reduction of the torsion-induced
helicity splitting.

Introducing the spinor
\begin{equation}
\Psi(\rho,\phi,z)=
\begin{pmatrix}
E_+(\rho,\phi,z)\\[2mm]
E_-(\rho,\phi,z)
\end{pmatrix},
\end{equation}
we may write Eq.~\eqref{eq:paraxial_components} compactly as
\begin{equation}
i\partial_z\Psi=\left[-\frac{1}{2k_0}\nabla_{\perp}^2\,\mathbb{I}+\Omega\rho\,\sigma_3\right]\Psi,
\label{eq:spinor_beam}
\end{equation}
where $\sigma_3$ acts in the circular basis. Equation~\eqref{eq:spinor_beam} is the effective
finite-width beam equation consistent with the local splitting $\Delta k_z \equiv k_z^{(-)}-k_z^{(+)}=2\Omega\rho$.

\subsection{Controlled approximations behind the beam equation}
\label{sec:controlled_approximations}

Equation~\eqref{eq:spinor_beam} should be understood as a minimal effective paraxial reduction of the local torsion-induced Maxwell dynamics rather than as a second exact field equation independently derived from the full Riemann--Cartan system. Its role is to embed, in the simplest finite-width setting, the same local helicity splitting obtained from the underlying covariant analysis.

Its validity rests on a controlled hierarchy of approximations. First, the beam is assumed to
propagate predominantly along the $z$ direction, so that the slowly varying envelope
approximation applies and higher longitudinal derivatives may be neglected relative to the carrier scale $k_0$. Second, the local circular propagation constants $k_z^{(\pm)}(\rho)$ obtained from the WKB-like analysis are promoted to weakly varying radial functions, which is appropriate when the transverse beam scale is large compared with the optical wavelength and the polarization state evolves adiabatically across the profile. Third, we retain only the leading helicity-diagonal torsion-induced phase splitting in the circular basis, while higher-order metric, connection, and polarization--spatial mode-mixing corrections are neglected. In
particular, any geometric connection terms associated with the transport of the local polarization basis across the transverse plane are neglected at this minimal level. Finally, the model assumes cylindrical symmetry and a purely radial torsional splitting, so that no additional azimuthal gauge structure is introduced at this stage.

Within this regime, the effective potentials $V_{\pm}(\rho)=k_0-k_z^{(\pm)}(\rho)$ capture the
leading torsion-dependent contribution to the paraxial beam evolution. The resulting
Eq.~\eqref{eq:spinor_beam} is therefore best viewed as the simplest finite-width transport model that remains explicitly anchored in the local geometric helicity splitting derived from
Eq.~\eqref{eq:exact_wave}. Precisely because of this restricted status, the model is useful: it isolates what follows robustly from torsional birefringence itself, without conflating those geometric consequences with additional material-specific ingredients such as absorption, disorder, microscopic inhomogeneity, non-geometric birefringence, or full vectorial basis-transport effects.

This immediately clarifies the model's physical content. If $\Omega=0$, the two circular sectors become degenerate, the spinor equation reduces to ordinary scalar paraxial diffraction, and the torsion-induced polarization textures vanish. If $\Omega\neq0$, the two helicity components accumulate distinct radial phases, and every finite-width beam acquires a spatially varying polarization transformation. What the minimal reduction predicts robustly is, therefore, beam-resolved polarization structuring; what it does not yet aim to capture is the full richness of realistic, defect-rich optical media.

Representative input states are Gaussian beams,
\begin{equation}
\Psi_G(\rho,\phi,0)=\mathcal{N}e^{-\rho^2/w_0^2}\chi_0,
\label{eq:gaussian_input}
\end{equation}
and vortex-like beams,
\begin{equation}
\Psi_{LG}(\rho,\phi,0)=\mathcal{N}\left(\frac{\rho}{w_0}\right)^{|m|}e^{-\rho^2/w_0^2}e^{im\phi}\chi_0,
\label{eq:lg_input}
\end{equation}
where $\chi_0$ is the input polarization spinor. For linearly polarized input, $\chi_0=(1,1)^T/\sqrt2$, the torsion-induced phase splitting manifests directly as a radial rotation of the local polarization axis.

A useful short-distance factorization is
\begin{equation}
\Psi(z)\approx e^{i\frac{z}{2k_0}\nabla_{\perp}^2}e^{-iz\Omega\rho\sigma_3}\Psi(0).
\label{eq:split_step_factorization}
\end{equation}
The first factor describes diffraction; the second contains the torsional phase. This decomposition makes explicit that all nontrivial polarization structuring in the minimal model
originates from the torsion-induced relative phase
\begin{equation}
U(\Omega,\rho,L)=e^{-i\Omega\rho L\sigma_3}.
\label{eq:torsion_unitary}
\end{equation}

\section{Stokes parameters and torsion-induced polarization textures}
\label{sec:textures}

The beam-level manifestation of torsion is most naturally described through the Stokes parameters. The circular-polarization basis is chosen as
\begin{equation}
\mathbf{e}_{\pm}=\frac{1}{\sqrt{2}}\left(\mathbf{e}_x \pm i\,\mathbf{e}_y\right),
\label{eq:circular_basis}
\end{equation}
so that the inverse transformation to the linear basis reads
\begin{equation}
E_x=\frac{E_++E_-}{\sqrt2},
\qquad
E_y=-\frac{E_+-E_-}{i\sqrt2}.
\label{eq:linear_basis}
\end{equation}
We then define
\begin{align}
S_0 &= |E_x|^2+|E_y|^2, \\
S_1 &= |E_x|^2-|E_y|^2, \\
S_2 &= 2\,\mathrm{Re}(E_xE_y^*), \\
S_3 &= 2\,\mathrm{Im}(E_xE_y^*).
\label{eq:stokes}
\end{align}

For an initially linearly polarized Gaussian beam and neglecting diffraction over short distances, the torsional propagation unitary in Eq.~\eqref{eq:torsion_unitary} yields
\begin{equation}
\Psi(\rho,\phi,z)=u_0(\rho,\phi)\frac{1}{\sqrt2}
\begin{pmatrix}
e^{-i\Omega\rho z}\\[1mm]
e^{+i\Omega\rho z}
\end{pmatrix}.
\label{eq:local_spinor_beam}
\end{equation}
Hence
\begin{align}
E_x(\rho,\phi,z)&=u_0(\rho,\phi)\cos(\Omega\rho z),\\
E_y(\rho,\phi,z)&=u_0(\rho,\phi)\sin(\Omega\rho z),
\label{eq:ExEy_torsion}
\end{align}
and the Stokes parameters become
\begin{align}
S_0 &= |u_0|^2,\\
S_1 &= |u_0|^2\cos\!\big[2\Omega\rho z\big],\\
S_2 &= |u_0|^2\sin\!\big[2\Omega\rho z\big],\\
S_3 &= 0.
\label{eq:stokes_torsion}
\end{align}
These expressions show explicitly how torsion enters the beam problem: the parameter $\Omega$
controls the radial oscillation of $S_1$ and $S_2$, and therefore the transverse polarization
texture. The local polarization angle is therefore
\begin{equation}
\psi(\rho,\phi,z)=\frac12\,\operatorname{arg}\!\left(S_1+iS_2\right)=\Omega\rho z,
\label{eq:psi_torsion}
\end{equation}
up to the usual branch convention, in direct agreement with the local rotation law
$\Delta\theta=\Omega\rho L$.

This is the main structured-light consequence of uniform torsion in the minimal model: the beam acquires a spatially varying polarization pattern because torsion lifts the helicity degeneracy differently at different radii. The observed polarization textures are therefore not an additional effect beyond torsion-induced optical activity; they are its finite-width manifestation.

\section{Geometric phase and restricted routes toward spin--orbit conversion}
\label{sec:soi}

Because torsion produces a relative phase between the two circular sectors, it is natural to
interpret the corresponding beam transformation as a geometric phase operation at the level of
polarization transport \cite{Pancharatnam1956,Cisowski2022}. At fixed radius, the polarization
state evolves according to the unitary in Eq.~\eqref{eq:torsion_unitary}, which acts as a rotation around the $z$ axis of the Poincar\'{e} sphere by angle $2\Omega\rho L$. In this sense, the torsional medium acts as an effective geometric-phase element whose action is governed by the combination $\Omega\rho L$. This interpretation is meant in the operational beam-optics sense relevant to polarization evolution, rather than as a full Berry-phase construction in an abstract parameter manifold.

However, one must distinguish carefully between radial polarization structuring and actual
spin--orbit conversion. In the minimal cylindrically symmetric model, the torsion-induced phase depends on $\rho$ but not on $\phi$. Therefore, it does not by itself impose a new azimuthal winding on the beam. The robust effect is a spatially varying polarization texture, not a necessary redistribution in orbital-angular-momentum sectors. This is consistent with the broader structured-light literature, where robust spin-to-orbital conversion generally requires an explicitly space-variant polarization element or an azimuthally structured geometric phase \cite{Marrucci2011,Padgett2017}.

A more general torsion-induced spin--orbit coupling would require an effective geometric connection with nontrivial azimuthal structure. Formally, if one performs a local unitary
transformation of the polarization basis,
\begin{equation}
\Psi=U(\rho,\phi,z)\widetilde\Psi,
\end{equation}
then the transformed equation contains the geometric gauge field
\begin{equation}
\mathbf{A}_g=iU^{\dagger}\nabla_{\perp}U.
\end{equation}
Only when $\mathbf{A}_g$ develops an azimuthal component can one expect robust OAM sidebands.
Thus, within the present minimal model derived from uniform torsion, polarization textures are
guaranteed, whereas spin--orbit conversion remains a possible but more restrictive extension.

\section{Beyond the minimal radial model: a torsional q-plate extension with azimuthal geometric connection}
\label{sec:qplate_extension}

The minimal cylindrically symmetric model developed in the previous sections makes clear that
uniform torsion generates a robust \emph{radial} helicity splitting,
\begin{equation}
\Delta k_z = 2\Omega \rho,
\end{equation}
and therefore a radius-dependent polarization rotation. At the same time, because the effective phase depends only on the radial coordinate, the model does not by itself impose a new azimuthal winding on the beam. This is precisely why the dominant structured-light signature of the minimal theory is polarization-texture formation rather than strong orbital-angular-momentum (OAM) conversion.

A natural question is whether one can preserve the torsion-induced radial splitting while
embedding it into a more general polarization structure that also carries nontrivial azimuthal
geometry. The simplest such extension is obtained by promoting the polarization basis to a
\emph{locally rotating frame} in the transverse plane, in close analogy with the action of a
$q$-plate or a space-variant Pancharatnam--Berry optical element \cite{Marrucci2011,Padgett2017,Cisowski2022}. In the present context, however, the point is not to replace the torsional splitting by an independent geometric-phase element, but to combine both in a single effective model. The resulting theory provides a controlled route beyond the minimal radial case and introduces an azimuthal geometric connection that can mediate torsion-assisted spin--orbit conversion.

\subsection{Locally rotating polarization basis}

Introducing the dimensionless variables $r=\rho/w_0$ and $\zeta=z/z_R$, with $z_R=k_0w_0^2$ as in Sec.~\ref{sec:numerics}, the minimal paraxial equation reads
\begin{equation}
i\partial_{\zeta}\Psi =
\left[
-\frac{1}{2}\nabla_r^2\,\mathbb{I}
+\Gamma_0 r\,\sigma_3
\right]\Psi,
\label{eq:minimal_dimensionless_recalled}
\end{equation}
with
\begin{equation}
\nabla_r^2 = \partial_r^2 + \frac{1}{r}\partial_r + \frac{1}{r^2}\partial_\phi^2.
\end{equation}
The term $\Gamma_0 r\,\sigma_3$ is the dimensionless representation of the torsion-induced
helicity splitting and remains the central physical ingredient of the theory.

To generate a nontrivial azimuthal geometric structure, we introduce a local unitary rotation of the polarization basis,
\begin{equation}
\Psi(r,\phi,\zeta) = U_q(\phi)\,\widetilde{\Psi}(r,\phi,\zeta),
\label{eq:qplate_basis_transform}
\end{equation}
where
\begin{equation}
U_q(\phi)=\exp\!\left[-\,i\,\frac{q\phi}{2}\sigma_2\right].
\label{eq:Uq_def}
\end{equation}
Here $q$ is a dimensionless topological charge controlling the azimuthal twisting of the local
polarization frame. In the simplest implementation, one may take $q\in\mathbb{Z}$, so that the
basis is single-valued under $\phi\to\phi+2\pi$. More general half-integer choices, as in the
standard $q$-plate literature, can also be considered once the physical equivalence of the local optic axis under a sign reversal is taken into account.

The operator $U_q(\phi)$ rotates the local polarization axis around the $\sigma_2$ direction
of the Poincar\'e sphere by angle $q\phi$. Accordingly, the fixed generator $\sigma_3$ is
mapped to the azimuthally varying operator
\begin{equation}
Q_q(\phi)
\equiv
U_q(\phi)\,\sigma_3\,U_q^\dagger(\phi)
=
\cos(q\phi)\,\sigma_3
+
\sin(q\phi)\,\sigma_1.
\label{eq:Qq_def}
\end{equation}
This expression is the key structural difference relative to the minimal theory: the torsional splitting axis is no longer globally fixed in polarization space, but winds azimuthally across the transverse plane.

With this definition, the most direct laboratory-frame generalization of the minimal torsional
model is
\begin{equation}
i\partial_{\zeta}\Psi
=
\left[
-\frac{1}{2}\nabla_r^2\,\mathbb{I}
+\Gamma_0 r\,Q_q(\phi)
\right]\Psi.
\label{eq:qplate_lab_frame}
\end{equation}
Equation~\eqref{eq:qplate_lab_frame} reduces to the previous minimal model when $q=0$, since
$Q_0=\sigma_3$. For $q\neq0$, however, the torsional splitting is embedded in a transverse
frame whose principal polarization axis rotates with $\phi$, thereby creating the possibility of azimuthal spin--orbit coupling.

\subsection{Co-rotating frame and geometric connection}

The same model admits a more revealing formulation in the co-rotating frame
$\widetilde{\Psi}$. Since $U_q$ depends only on the transverse coordinate $\phi$, we have
\begin{equation}
\nabla_\perp \Psi
=
\nabla_\perp(U_q\widetilde{\Psi})
=
U_q\left(\nabla_\perp - i\mathbf{A}_g\right)\widetilde{\Psi},
\end{equation}
where the geometric gauge field is
\begin{equation}
\mathbf{A}_g
=
i\,U_q^\dagger\nabla_\perp U_q.
\label{eq:Ag_general}
\end{equation}
Using Eq.~\eqref{eq:Uq_def} and the polar-coordinate identity
$\nabla_\perp = \hat{\mathbf r}\,\partial_r + \hat{\boldsymbol\phi}\,r^{-1}\partial_\phi$, we
find
\begin{equation}
\partial_\phi U_q
=
-\,i\,\frac{q}{2}\sigma_2\,U_q,
\end{equation}
and therefore
\begin{equation}
\mathbf{A}_g
=
\hat{\boldsymbol\phi}\,\frac{q}{2r}\,\sigma_2.
\label{eq:Ag_qplate}
\end{equation}
This is the explicit azimuthal geometric connection anticipated in Sec.~\ref{sec:soi}. Its appearance is the central mathematical reason why the present extension can support nontrivial OAM conversion.

Substituting Eq.~\eqref{eq:qplate_basis_transform} into Eq.~\eqref{eq:qplate_lab_frame}, multiplying on the left by $U_q^\dagger$, and using Eq.~\eqref{eq:Qq_def}, we obtain the co-rotating-frame equation
\begin{equation}
i\partial_{\zeta}\widetilde{\Psi}
=
\left[
-\frac{1}{2}\left(\nabla_\perp-i\mathbf{A}_g\right)^2
+\Gamma_0 r\,\sigma_3
\right]\widetilde{\Psi}.
\label{eq:qplate_corotating}
\end{equation}
Equation~\eqref{eq:qplate_corotating} is the natural torsional $q$-plate generalization of the
minimal beam model. It preserves the same torsion-driven radial splitting $\Gamma_0 r\sigma_3$, but now it is supplemented by a nontrivial covariant derivative in the transverse plane.

Because $A_{g,r}=0$ and $A_{g,\phi}=q(2r)^{-1}\sigma_2$, Eq.~\eqref{eq:qplate_corotating}
can be written explicitly as
\begin{align}
i\partial_{\zeta}\widetilde{\Psi}
=
\Bigg[
&-\frac{1}{2}
\left(
\partial_r^2
+\frac{1}{r}\partial_r
+\frac{1}{r^2}
\left(
\partial_\phi
-i\frac{q}{2}\sigma_2
\right)^2
\right)\notag\\&
+\Gamma_0 r\,\sigma_3
\Bigg]\widetilde{\Psi}.
\label{eq:qplate_explicit}
\end{align}
For $r>0$, the angular part may be expanded as
\begin{equation}
\left(
\partial_\phi
-i\frac{q}{2}\sigma_2
\right)^2
=
\partial_\phi^2
-
iq\,\sigma_2\,\partial_\phi
-
\frac{q^2}{4},
\end{equation}
so that
\begin{align}
i\partial_{\zeta}\widetilde{\Psi}
=
\Bigg[
&-\frac{1}{2}
\left(
\partial_r^2+\frac{1}{r}\partial_r+\frac{1}{r^2}\partial_\phi^2
\right)\mathbb{I}
+\frac{i q}{2r^2}\sigma_2\,\partial_\phi
+\frac{q^2}{8r^2}\mathbb{I}
\notag\\
&\hspace{3.2cm}
+\Gamma_0 r\,\sigma_3
\Bigg]\widetilde{\Psi}.
\label{eq:qplate_expanded}
\end{align}
This expanded form makes the new physics especially transparent. The term $\frac{i q}{2r^2}\sigma_2\partial_\phi$ explicitly couples polarization and orbital structure, while the scalar term $q^2/(8r^2)$ acts as an effective centrifugal correction generated by the azimuthally rotating polarization frame.

A useful conceptual point is that $\mathbf{A}_g$ by itself does \emph{not} introduce a new
physical interaction when $\Gamma_0=0$: in that limit the model is gauge-equivalent to the free paraxial equation, since the geometric connection arises purely from a local change of basis. Observable new effects appear only when the torsion-induced splitting $\Gamma_0 r\sigma_3$ is present simultaneously with the azimuthal frame winding. It is the \emph{combination} of torsional birefringence and azimuthal geometric structure that produces genuine spin--orbit conversion.

\subsection{Selection rules and perturbative OAM sidebands}

The azimuthal content of the extended model can be read directly from the laboratory-frame operator $Q_q(\phi)$. Using
\begin{align}
\cos(q\phi) &= \frac{1}{2}\left(e^{iq\phi}+e^{-iq\phi}\right),\\
\sin(q\phi) &= \frac{1}{2i}\left(e^{iq\phi}-e^{-iq\phi}\right),
\end{align}
Eq.~\eqref{eq:Qq_def} becomes
\begin{equation}
Q_q(\phi)
=
\frac{1}{2}e^{iq\phi}\left(\sigma_3-i\sigma_1\right)
+
\frac{1}{2}e^{-iq\phi}\left(\sigma_3+i\sigma_1\right).
\label{eq:Qq_fourier}
\end{equation}
This immediately implies the selection rule
\begin{equation}
m \longrightarrow m\pm q,
\label{eq:selection_rule}
\end{equation}
for the azimuthal harmonics of the field.

To see this more explicitly, write the beam in angular Fourier components,
\begin{equation}
\Psi(r,\phi,\zeta)
=
\sum_{m=-\infty}^{\infty}
\psi_m(r,\zeta)\,e^{im\phi}.
\label{eq:angular_expansion_qplate}
\end{equation}
Then the torsional interaction term in Eq.~\eqref{eq:qplate_lab_frame} generates matrix elements
\begin{align}
\mel{m'}{\Gamma_0 r\,Q_q(\phi)}{m}
&=
\frac{\Gamma_0 r}{2}\left(\sigma_3-i\sigma_1\right)\delta_{m',m+q}
\notag\\&+
\frac{\Gamma_0 r}{2}\left(\sigma_3+i\sigma_1\right)\delta_{m',m-q}.
\label{eq:matrix_element_selection}
\end{align}
Hence, unlike the purely radial minimal model, the extended Hamiltonian mixes different OAM sectors already at first order in the torsional coupling.

The short-distance behavior makes this especially transparent. Neglecting diffraction over a small propagation interval $\zeta$, one has
\begin{equation}
\Psi_{\rm out}(r,\phi,\zeta)
\approx
\left[
\mathbb{I}
-
i\,\Gamma_0 r\zeta\,Q_q(\phi)
\right]\Psi_{\rm in}(r,\phi,0).
\label{eq:short_distance_qplate}
\end{equation}
For an input Gaussian beam with no initial azimuthal winding,
\begin{equation}
\Psi_{\rm in}(r,\phi,0)
=
u_0(r)\,\chi_0,
\label{eq:qplate_gaussian_input}
\end{equation}
Eq.~\eqref{eq:Qq_fourier} gives
\begin{align}
\Psi_{\rm out}
\approx
u_0(r)\chi_0
-\,
i\,\frac{\Gamma_0 r\zeta}{2}\nu_0(r)
\Big[
&e^{iq\phi}\left(\sigma_3-i\sigma_1\right)\chi_0
\notag\\
&+
e^{-iq\phi}\left(\sigma_3+i\sigma_1\right)\chi_0
\Big].
\label{eq:first_order_sidebands}
\end{align}
Thus, even when the input beam carries only the $m=0$ sector, the output field acquires
$m=\pm q$ sidebands at first order in $\Gamma_0\zeta$. The corresponding sideband powers
therefore scale as
\begin{equation}
P_{\pm q}(\zeta)\propto (\Gamma_0\zeta)^2
\int_0^\infty |u_0(r)|^2 r^3\,dr,
\label{eq:sideband_scaling}
\end{equation}
up to polarization-dependent prefactors determined by the spinor $\chi_0$. This quadratic
growth in the sideband power is the natural perturbative signature of the torsion-assisted
spin--orbit conversion channel opened by the azimuthal connection.

The same conclusion can be stated in the co-rotating frame. There, the term $\frac{i q}{2r^2}\sigma_2\partial_\phi$ in Eq.~\eqref{eq:qplate_expanded} couples the orbital generator $-i\partial_\phi$ to polarization through $\sigma_2$, while the radial torsional splitting keeps the two circular sectors spectrally distinct. The extended model thus combines two
ingredients: a \emph{spin-dependent radial phase accumulation} inherited from the torsional medium and a \emph{nontrivial azimuthal geometric connection} inherited from the rotating polarization frame. Their coexistence enables robust OAM redistribution.

\begin{figure*}[t]
\centering
\includegraphics[width=\textwidth]{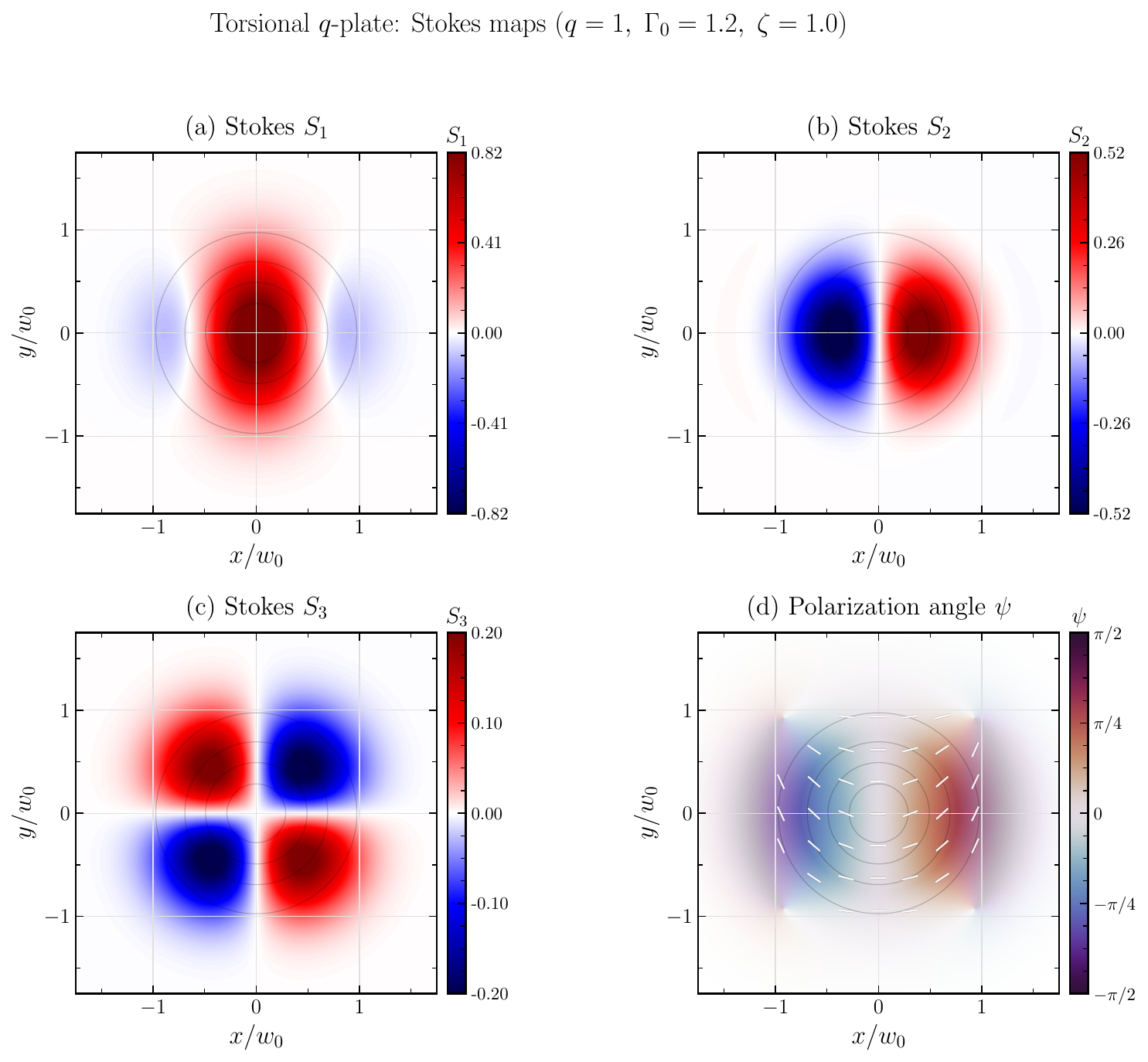}
\caption{Local Stokes maps for the torsional $q$-plate extension in the short-distance, diffraction-neglected regime, for a Gaussian input beam with homogeneous linear polarization, $q=1$, $\Gamma_0=1.2$, and $\zeta=1.0$. Panel (a) shows $S_1(x/w_0,y/w_0)$, panel (b) shows $S_2(x/w_0,y/w_0)$, panel (c) shows $S_3(x/w_0,y/w_0)$, and panel (d) shows the local polarization angle $\psi=\frac{1}{2}\arg(S_1+iS_2)$, with polarization-direction arrows overlaid. Unlike the minimal radial model, in which the torsion-induced texture depends only on the radial coordinate, the present extension displays explicit azimuthal modulation generated by the locally rotating polarization frame $U_q(\phi)$. The appearance of a nontrivial $S_3$ pattern indicates that the beam acquires local ellipticity, while the angular structure of $\psi$ shows that the polarization rotation is no longer purely radial. This figure, therefore, provides the real-space signature of the azimuthal geometric connection that enables torsion-assisted spin--orbit conversion.}
\label{fig:qplate_torsion_maps}
\end{figure*}

Figure~\ref{fig:qplate_torsion_maps} presents the first real-space signature of the torsional $q$-plate extension introduced above. In contrast with the minimal uniform-torsion model, where the local phase depends only on the radial coordinate and therefore generates purely annular polarization textures, the present extension combines the same radial torsional splitting with a polarization basis that rotates azimuthally across the transverse plane. The resulting field is therefore no longer characterized by radial structuring alone: the Stokes parameters acquire explicit $\phi$-dependent modulation, and the output beam develops a genuinely mixed radial and azimuthal polarization texture.

Figure \ref{fig:qplate_torsion_maps}(a) shows the map of $S_1$, which is no longer organized into concentric annular domains. Instead, the linear-polarization contrast becomes anisotropic across the beam cross-section, indicating that the preferred local polarization axis now depends on both $r$ and $\phi$. Figure \ref{fig:qplate_torsion_maps}(b) displays the corresponding map of $S_2$, which exhibits a complementary angular structure shifted relative to $S_1$. Together, these two panels show that the pair $(S_1,S_2)$ no longer represents a purely radial rotation of a homogeneous linear state, but rather an azimuthally structured redistribution of the linear-polarization sector induced by the operator
\begin{equation}
  Q_q(\phi)=\cos(q\phi)\sigma_3+\sin(q\phi)\sigma_1.  
\end{equation}
This is precisely the real-space manifestation of embedding the torsion-induced splitting into a locally rotating polarization frame.

The most distinctive new feature appears in Fig. \ref{fig:qplate_torsion_maps}(c), which displays the Stokes parameter $S_3$. In the minimal radial model, $S_3=0$ identically for the linearly polarized Gaussian input considered in the short-distance regime, so the field remains locally linear even though its orientation varies with radius. Here, by contrast, the $q$-dependent transverse frame generates a nontrivial quadrupolar $S_3$ pattern, showing that the output beam acquires local ellipticity. This is an important qualitative change: the extended model does not merely rotate the local linear-polarization axis, but mixes the polarization components in such a way that the field samples different regions of the Poincar\'e sphere across the beam profile.

Figure \ref{fig:qplate_torsion_maps}(d) summarizes the same effect in terms of the local polarization angle $\psi=\frac{1}{2}\arg(S_1+iS_2)$. In the minimal torsional model, $\psi$ obeys the simple radial law $\psi=\Gamma_0 r\zeta$, up to the usual branch convention. In the present extension, however, the angle map becomes azimuthally modulated and is no longer a function of radius alone. The overlaid polarization-direction arrows make this especially clear: the local director field now winds across the beam in a manner controlled jointly by the radial torsional phase and the topological charge $q$ of the rotating basis. The figure, therefore, provides a direct visual diagnostic of the new geometric ingredient introduced by the extension, namely the azimuthal connection
\begin{equation}
\mathbf A_g=\hat{\boldsymbol\phi}\,\frac{q}{2r}\sigma_2,
\end{equation}
whose presence opens the route toward robust orbital-angular-momentum sidebands.

Taken together, the four panels of Fig.~\ref{fig:qplate_torsion_maps} establish the physical role of the torsional $q$-plate model before any spectral decomposition is performed. The beam no longer exhibits only radial-polarization structuring; instead, it develops a coupled radial and azimuthal texture, together with local ellipticity, as the real-space precursor to the spin--orbit conversion channels analyzed below.

\begin{figure*}[t]
\centering
\includegraphics[width=\textwidth]{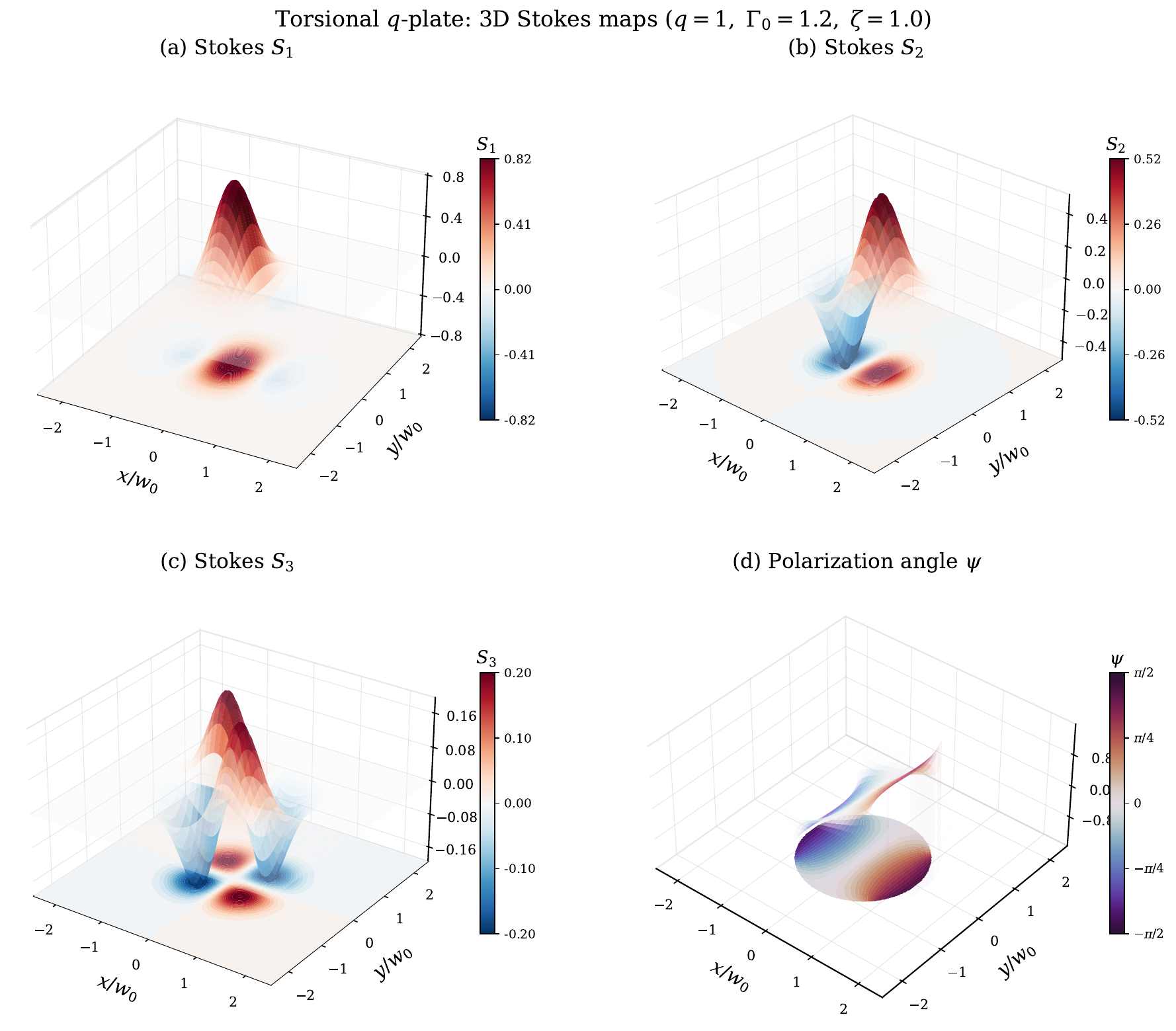}
\caption{Three-dimensional visualization of the local Stokes maps for the torsional $q$-plate extension in the short-distance, diffraction-neglected regime, for a Gaussian input beam with homogeneous linear polarization, $q=1$, $\Gamma_0=1.2$, and $\zeta=1.0$. Panel (a) shows $S_1(x/w_0,y/w_0)$, panel (b) shows $S_2(x/w_0,y/w_0)$, panel (c) shows $S_3(x/w_0,y/w_0)$, and panel (d) shows the local polarization angle $\psi=\frac{1}{2}\arg(S_1+iS_2)$. The floor projections reproduce the corresponding transverse maps, while the three-dimensional surfaces emphasize the azimuthally modulated structure induced by the locally rotating polarization frame. In particular, the quadrupolar structure of $S_3$ becomes especially transparent in this representation. The apparent discontinuities in panel (d) are due to the principal-branch representation of the angle variable $\psi$, rather than to any additional physical singularity.}
\label{fig:qplate_torsion_maps_3d}
\end{figure*}

Figure~\ref{fig:qplate_torsion_maps_3d} provides a complementary three-dimensional rendering of the same representative configuration shown in Fig.~\ref{fig:qplate_torsion_maps}. Its purpose is not to introduce new observables, but to make the morphology of the torsion-induced $q$-plate textures more visually transparent. In the two-dimensional maps, the physical content is already clear: the azimuthally rotating polarization frame breaks the purely radial structure of the minimal model and generates a mixed radial--azimuthal modulation of the Stokes sector. The present three-dimensional view reinforces this conclusion by displaying the same quantities as surfaces over the transverse plane.

Figures \ref{fig:qplate_torsion_maps_3d}(a) and \ref{fig:qplate_torsion_maps_3d}(b) show the surfaces associated with $S_1$ and $S_2$. In this representation, the anisotropic redistribution of the linear-polarization sector becomes especially evident. Rather than forming concentric annular domains, as in the minimal radial theory, the two Stokes components now develop a directional structure that depends simultaneously on $r$ and $\phi$.
The three-dimensional surfaces, therefore, provide a more geometric visualization of how the operator
\[
Q_q(\phi)=\cos(q\phi)\sigma_3+\sin(q\phi)\sigma_1
\]
reorients the local polarization response across the beam profile.

Figure \ref{fig:qplate_torsion_maps_3d}(c), showing $S_3$, is particularly informative. In the minimal model and for the same class of linearly polarized Gaussian inputs, $S_3$ vanishes identically in the short-distance regime, so the field remains locally linear even when its orientation changes with radius. In the torsional $q$-plate extension, by contrast, the field acquires a nontrivial quadrupolar ellipticity pattern. This feature is already visible in the two-dimensional map of Fig.~\ref{fig:qplate_torsion_maps}, but the three-dimensional rendering highlights it much more clearly, making explicit that the new extension drives the beam away from purely linear local polarization states and into a spatially varying exploration of the Poincar\'e sphere.

Figure \ref{fig:qplate_torsion_maps_3d}(d) shows the local polarization angle
\[
\psi=\frac{1}{2}\arg(S_1+iS_2).
\]
As in the two-dimensional representation, the angle is displayed on its principal branch and
therefore exhibits the expected wrapping discontinuities. In the three-dimensional surface, these appear as sharp transitions, but they should not be interpreted as additional physical structures: they are simply the standard consequence of representing an angular variable modulo $\pi$. Physically, the important point is that $\psi$ is no longer a function of radius alone. The surface makes clear that the local director field now carries genuine azimuthal modulation, controlled jointly by the radial torsional phase accumulation and the topological charge $q$ of the rotating basis.

Taken together, Figs.~\ref{fig:qplate_torsion_maps} and \ref{fig:qplate_torsion_maps_3d} provide a complete real-space picture of the torsional $q$-plate extension. The former gives the most direct and compact view of the local Stokes maps, while the latter emphasizes their geometric morphology and makes the emergence of anisotropy and local ellipticity more visually explicit. This prepares the ground for the next step, namely the spectral characterization of the same extension in orbital-angular-momentum space.

While Figs.~\ref{fig:qplate_torsion_maps} and \ref{fig:qplate_torsion_maps_3d} establish the
real-space structure of the torsional $q$-plate extension, they do not yet quantify how the
azimuthal content of the beam is redistributed among orbital harmonics. That information is carried by the OAM weights $P_m$, to which we now turn.

\begin{figure*}[t]
\centering
\includegraphics[width=\textwidth]{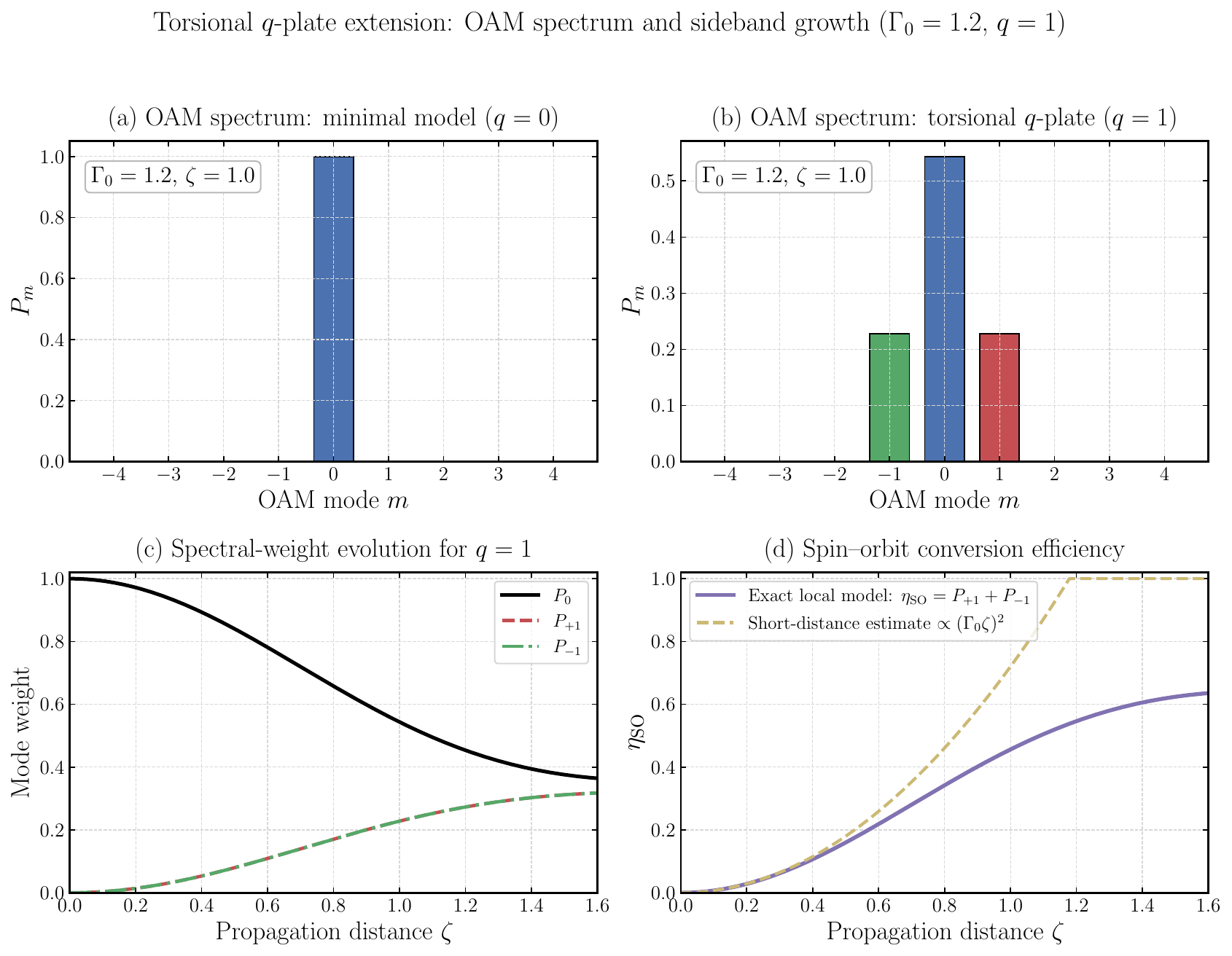}
\caption{Orbital-angular-momentum content in the torsional $q$-plate extension for a Gaussian input beam with homogeneous linear polarization. Panel (a) shows the OAM spectrum $P_m$ for the minimal radial model ($q=0$), where the torsion-induced phase depends only on the radial coordinate and the azimuthal content remains concentrated in the initial mode. Panel (b) shows the corresponding spectrum for the torsional $q$-plate extension with $q=1$, for which symmetry-allowed sidebands $m\rightarrow m\pm q$ appear. Panel (c) displays the evolution of the dominant spectral weights $P_0$, $P_{+1}$, and $P_{-1}$ as functions of $\zeta$, showing the transfer of weight from the initial azimuthal sector into the sidebands. Panel (d) shows the spin--orbit conversion efficiency $\eta_{\rm SO}=P_{+1}+P_{-1}$, together with the short-distance perturbative guide $\eta_{\rm SO}\propto (\Gamma_0\zeta)^2$. The figure demonstrates that the azimuthal geometric connection activates a genuine spin--orbit conversion channel absent in the purely radial minimal theory.}
\label{fig:qplate_oam_spectrum}
\end{figure*}

Figure~\ref{fig:qplate_oam_spectrum} provides the OAM-space counterpart of the real-space textures shown in Figs.~\ref{fig:qplate_torsion_maps} and \ref{fig:qplate_torsion_maps_3d}. In the minimal radial model, the torsion-induced phase depends only on the radial coordinate, so although the beam develops nontrivial polarization textures, its azimuthal content remains essentially unchanged. In the present extension, by contrast, the azimuthally rotating polarization frame introduces the operator
\[
Q_q(\phi)=\cos(q\phi)\sigma_3+\sin(q\phi)\sigma_1,
\]
whose Fourier decomposition contains the harmonics $e^{\pm iq\phi}$. As a result, the field is no longer confined to its initial azimuthal sector, and the OAM spectrum becomes a primary observable.

Figure \ref{fig:qplate_oam_spectrum}(a) displays the reference spectrum for the minimal radial model, corresponding to $q=0$. As expected, the distribution remains concentrated in the initial mode $m=0$. This explicitly confirms that the radial torsional splitting alone does not generate strong azimuthal mode conversion. The beam may acquire substantial polarization structuring in real space, but its OAM content remains essentially trivial. In this sense, the minimal theory acts primarily as a spatially resolved polarization rotator rather than as an efficient spin--orbit converter.

Figure \ref{fig:qplate_oam_spectrum}(b) shows the corresponding spectrum for the torsional $q$-plate extension with $q=1$. The main new feature is the appearance of sidebands in the neighboring azimuthal sectors $m=\pm1$. These are precisely the symmetry-allowed channels predicted by the selection rule
\[
m\rightarrow m\pm q,
\]
derived from the Fourier decomposition of
\[
Q_q(\phi)=\cos(q\phi)\sigma_3+\sin(q\phi)\sigma_1.
\]
Thus, once the radial torsional splitting is embedded in a locally rotating polarization frame, the beam no longer remains confined to its initial azimuthal sector: part of its spectral weight is transferred into the sidebands by the combined action of the torsional phase accumulation and the azimuthal geometric connection.

Figure \ref{fig:qplate_oam_spectrum}(c) clarifies this transfer by showing the evolution of the dominant-mode weights $P_0$, $P_{+1}$, and $P_{-1}$ as functions of propagation distance. Starting from an input beam concentrated in $m=0$, the extended model shows a progressive decrease of $P_0$ accompanied by a symmetric growth of $P_{+1}$ and $P_{-1}$. The near equality of the two sideband curves reflects the symmetry of the present $q=1$ extension for a linearly polarized Gaussian input. The panel therefore provides a direct dynamical visualization of the spectral weight transfer induced by the torsion-assisted spin--orbit channel.

Figure \ref{fig:qplate_oam_spectrum}(d) summarizes the same effect in terms of the spin--orbit conversion efficiency
\[
\eta_{\rm SO}=P_{+1}+P_{-1}.
\]
The exact local model shows a monotonic increase of $\eta_{\rm SO}$ with propagation distance,
indicating that the sideband channels become progressively more populated as the torsional phase is accumulated. The dashed curve gives the short-distance perturbative guide
$\eta_{\rm SO}\propto(\Gamma_0\zeta)^2$, which captures the expected initial quadratic growth of the sideband power. At larger $\zeta$, deviations naturally appear, since the perturbative
expression is valid only in the weak-conversion regime, whereas the exact local model already
includes the full nonperturbative mixing generated by the operator $J_q$.

Taken together, the four panels of Fig.~\ref{fig:qplate_oam_spectrum} establish the central
distinction between the minimal and extended theories. In the former, torsion generates robust
polarization textures without significant OAM redistribution. In the latter, the same torsional splitting is embedded in a rotating transverse polarization frame, and the beam develops symmetry-allowed sidebands in orbital-angular-momentum space. Figure~\ref{fig:qplate_oam_spectrum}, therefore, confirms that the azimuthal geometric connection opens a qualitatively new channel of structured-light dynamics, transforming the uniform-torsion beam model from a purely radial polarization-structuring mechanism into a genuine torsion-assisted spin--orbit converter.

\begin{figure*}[t]
    \centering
    \includegraphics[width=\textwidth]{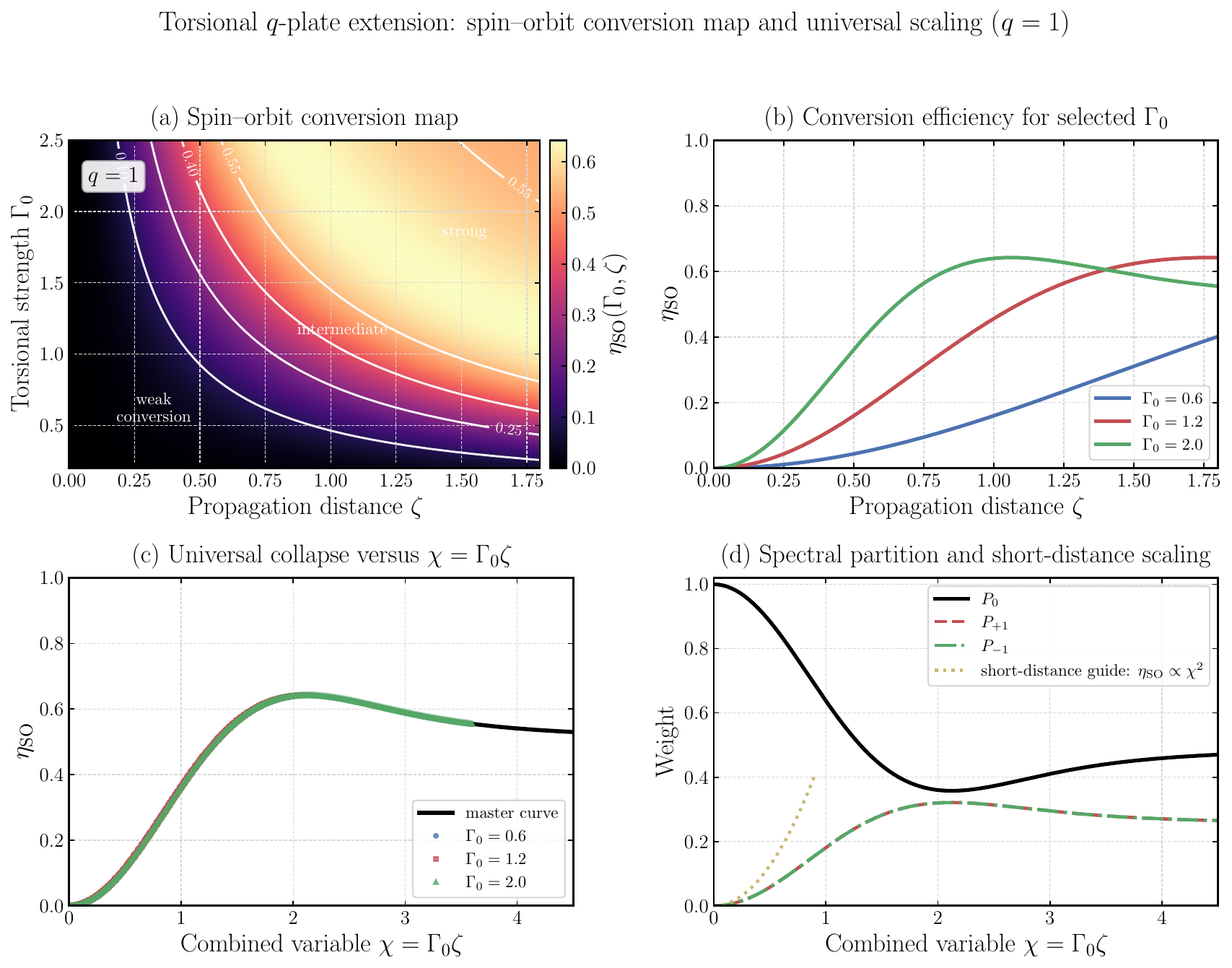}
    \caption{
    Spin--orbit conversion map for the torsional $q$-plate extension in the short-distance,
    diffraction-neglected regime, for $q=1$. Panel (a) shows the conversion efficiency     $\eta_{\rm SO}=P_{+1}+P_{-1}$ in the $(\Gamma_0,\zeta)$ plane, with white contours marking
    representative efficiency levels. Panel (b) displays $\eta_{\rm SO}(\zeta)$ for selected values of the torsional strength $\Gamma_0$, showing how the conversion sets in more rapidly as $\Gamma_0$ increases. Panel (c) presents the universal collapse of the same data when plotted against the combined variable $\chi=\Gamma_0\zeta$, demonstrating that the local conversion dynamics are controlled primarily by this product. Panel (d) shows the corresponding spectral partition into the carrier and sideband sectors, $P_0$, $P_{+1}$, and $P_{-1}$, together with the short-distance perturbative guide
    $\eta_{\rm SO}\propto \chi^2$. The figure shows that, within the local torsional $q$-plate model, the onset and strength of spin--orbit conversion obey a simple scaling law and can be summarized by a compact operating diagram in the $(\Gamma_0,\zeta)$ plane.
    } \label{fig:qplate_conversion_map}
\end{figure*}

While Fig.~\ref{fig:qplate_oam_spectrum} identifies the spectral sidebands generated by the
torsional $q$-plate extension, it is also useful to summarize the same process in a compact
regime-space representation. This is done in Fig.~\ref{fig:qplate_conversion_map}, which shows how the spin--orbit conversion efficiency depends jointly on the torsional strength and the propagation distance, and reveals the scaling structure of the local model.

Figure~\ref{fig:qplate_conversion_map} summarizes the spin--orbit conversion dynamics of the
torsional $q$-plate extension in a compact operational form. While Fig.~\ref{fig:qplate_oam_spectrum} established the appearance of the symmetry-allowed sidebands $m=\pm1$ and their growth with propagation distance, the present figure reorganizes the same physics into a regime map and shows that, within the local diffraction-neglected model, the conversion process is governed predominantly by the combined variable
\[
\chi=\Gamma_0\zeta.
\]
This makes the new extension especially transparent: the azimuthal geometric connection not only opens the sideband channels, but does so according to a simple scaling structure.

Figure \ref{fig:qplate_conversion_map}(a) shows the spin--orbit conversion efficiency
\[
\eta_{\rm SO}=P_{+1}+P_{-1}
\]
in the $(\Gamma_0,\zeta)$ plane. The map reveals a smooth crossover from a weak conversion
region at small $\Gamma_0$ and short propagation distance to an intermediate and then stronger
conversion regime as the accumulated torsional phase increases. The contour lines are approximately hyperbolic, reflecting the fact that the dominant control parameter is the product $\Gamma_0\zeta$ rather than $\Gamma_0$ or $\zeta$ separately. In this sense, the panel
provides the spin--orbit analogue of the regime maps previously used to characterize polarization texturing in the minimal radial model.

Figure \ref{fig:qplate_conversion_map}(b) displays representative cuts $\eta_{\rm SO}(\zeta)$ for selected values of $\Gamma_0$. As expected, greater torsional strength shifts the conversion onset to shorter propagation distances and leads to higher efficiencies within the same $\zeta$ window. The curves are not strictly linear or purely saturating: in the exact local model, the conversion is the result of the full radial average of oscillatory torsional phase factors, so different values of $\Gamma_0$ may display slightly different degrees of buildup and partial redistribution as $\zeta$ increases. This is especially visible for the largest $\Gamma_0$ shown, where the curve rises rapidly and then exhibits a mild turnover. Such behavior is not a numerical artifact, but a direct consequence of the exact local mixing encoded in the model.

The central scaling result is displayed in Fig. \ref{fig:qplate_conversion_map}(c). When the same data are replotted as functions of
\[
\chi=\Gamma_0\zeta,
\]
the different curves collapse onto a single master curve. This demonstrates that, in the present local approximation, the spin--orbit conversion dynamics is controlled by a single effective parameter combining torsional strength and propagation distance. The collapse is physically important because it shows that the conversion map in Fig. \ref{fig:qplate_conversion_map}(a) is not merely a numerical survey, but the manifestation of an underlying scaling law. It also gives the section a more universal character: the results are not tied to one particular choice of $\Gamma_0$ and $\zeta$, but follow from their product.

Figure \ref{fig:qplate_conversion_map}(d) complements this picture by showing the associated spectral partition among the dominant azimuthal sectors. The carrier contribution $P_0$ decreases as $\chi$ grows, while the two sidebands $P_{+1}$ and $P_{-1}$ increase symmetrically, as expected for the $q=1$ extension with a linearly polarized Gaussian input. The symmetry between $P_{+1}$ and $P_{-1}$ confirms that the local torsional $q$-plate does not favor one of the two neighboring OAM channels over the other in the present configuration. The same panel also includes the short-distance perturbative guide
\[
\eta_{\rm SO}\propto \chi^2,
\]
which captures the initial quadratic growth of the conversion efficiency. At larger $\chi$,
deviations naturally appear, since the perturbative expression is valid only in the weak-conversion regime, whereas the exact local model already includes the full nonperturbative redistribution among the populated azimuthal sectors.

Taken together, the four panels of Fig.~\ref{fig:qplate_conversion_map} elevate the new section from a purely qualitative extension to a compact quantitative framework. The azimuthal geometric connection introduced by the torsional $q$-plate model not only generates sidebands in OAM space, as shown in Fig.~\ref{fig:qplate_oam_spectrum}, but does so according to a simple regime structure and a universal scaling law. The figure therefore provides the most operational summary of the new physics introduced beyond the minimal radial model: a controlled torsion-assisted spin--orbit conversion process governed, at the local level, by the combined phase variable $\chi$.

\subsection{Relation to the minimal model and expected observables}

Several limiting cases help clarify the structure of the extension. If $q=0$, then $U_q=\mathbb{I}$, $\mathbf{A}_g=0$, $Q_q=\sigma_3$, and the theory reduces exactly to the
minimal radial model of Secs.~\ref{sec:paraxial}--\ref{sec:results}. If $\Gamma_0=0$, the
torsional splitting disappears and the apparent connection $\mathbf{A}_g$ is a pure gauge
artifact of the chosen local basis, so no physical spin--orbit conversion remains. Only the
combined regime $q\neq0$ and $\Gamma_0\neq0$ represents a genuinely new optical system.

The extended model suggests a qualitatively richer observable set than the one studied in the
minimal case. First, the OAM weights $P_m(\zeta)$, introduced previously mainly as consistency
checks, become central dynamical observables. In particular, the growth of $P_{\pm q}$ from an
initially $m=0$ beam provides a direct signature of torsion-assisted spin--orbit conversion.
Second, the Stokes maps are expected to acquire genuine $\phi$-dependent structure, including
azimuthally modulated patterns in $S_1$ and $S_2$, and, depending on the input polarization
and propagation distance, a nontrivial $S_3$ component may also emerge locally. Third, the
global observables $\bar{\theta}$ and $\mathcal{C}$ remain meaningful, but they no longer
fully characterize the field by themselves, because the beam can now develop angular as well as radial polarization complexity.

A practical numerical implementation of Eq.~\eqref{eq:qplate_lab_frame} or Eq.~\eqref{eq:qplate_corotating} would therefore require a genuinely two-dimensional paraxial
solver, either in Cartesian $(x,y)$ coordinates or on a polar grid in $(r,\phi)$. From the
conceptual standpoint of the present article, however, the key point is already visible at the
analytic level: once the torsional splitting is embedded in a rotating transverse polarization
frame, the theory acquires an azimuthal geometric connection, and the previously absent route
toward robust OAM sidebands becomes available.

Finally, one should note that the idealized connection in Eq.~\eqref{eq:Ag_qplate} contains the usual $1/r$ singularity at the origin familiar from $q$-plate-like models. In realistic
implementations, the beam core or the material response regularizes this singular behavior. A more refined version of the model could therefore replace $q/(2r)$ by $q f(r)/(2r)$, where
$f(r)\to0$ as $r\to0$ and $f(r)\to1$ outside the core. Such a regularization does not alter
the main selection rule in Eq.~\eqref{eq:selection_rule}; it only smooths the behavior near the origin and may be useful in future numerical work.

In this sense, the torsional $q$-plate extension proposed here supplies the missing conceptual
bridge between the beam-resolved polarization textures of the minimal uniform-torsion model and a more general class of torsion-assisted spin--orbit structured-light effects.

\section{Numerical framework and torsion-controlled observables}
\label{sec:numerics}

For numerical work, it is convenient to introduce the dimensionless variables
\begin{equation}
r=\frac{\rho}{w_0},
\qquad
\zeta=\frac{z}{z_R},
\qquad
z_R=k_0w_0^2.
\end{equation}
The beam equation then becomes
\begin{equation}
i\partial_{\zeta}\Psi=\left[-\frac12\nabla_r^2\,\mathbb{I}+\Gamma(r)\sigma_3\right]\Psi,
\label{eq:dimensionless_eq}
\end{equation}
with
\begin{equation}
\Gamma(r)=z_R\Omega w_0 r \equiv \Gamma_0 r.
\label{eq:Gamma_linear_torsion}
\end{equation}
The key point is that $\Gamma_0$ is not an arbitrary fitting constant: it is the dimensionless
form of the torsion parameter. Setting $\Gamma_0=0$ therefore corresponds exactly to turning off
torsion.

The primary observables are the transverse maps of $S_0,S_1,S_2$, the polarization angle
$\psi$, and the azimuthal decomposition weights
\begin{equation}
P_m(z)=\int_0^{\infty}|c_m(\rho,z)|^2\rho\,d\rho,
\end{equation}
where
\begin{equation}
c_m(\rho,z)=\frac{1}{2\pi}\int_0^{2\pi}E(\rho,\phi,z)e^{-im\phi}d\phi.
\end{equation}

To characterize the net output polarization in a way that remains meaningful under radial
dephasing, we define the spatially integrated Stokes components
\begin{equation}
\bar S_j(z)=
\frac{\int S_j(\rho,\phi,z)\,\rho\,d\rho\,d\phi}
{\int S_0(\rho,\phi,z)\,\rho\,d\rho\,d\phi},
\qquad j=1,2,
\label{eq:mean_stokes}
\end{equation}
and the corresponding beam-averaged polarization angle
\begin{equation}
\bar{\theta}(z)=\frac12\,\operatorname{arg}\!\left(\bar S_1(z)+i\,\bar S_2(z)\right).
\label{eq:theta_bar}
\end{equation}
This quantity measures the orientation of the \emph{net} linear polarization extracted from the integrated Stokes vector. Unlike a direct average of the local angle $\psi$, it remains well posed when the local polarization rotates through several branches across the beam profile. It should be interpreted as a global output angle rather than a strictly cumulative phase variable.

A complementary scalar diagnostic of texture formation is obtained from the spatial variation of the normalized linear-polarization field
\begin{equation}
s_1(\rho,\phi,z)=\frac{S_1}{S_0},
\qquad
s_2(\rho,\phi,z)=\frac{S_2}{S_0},
\end{equation}
through the inhomogeneity measure
\begin{equation}
\mathcal{C}(z)=
\left[
\frac{\int \left[(s_1-\bar S_1)^2+(s_2-\bar S_2)^2\right]S_0\,
\rho\,d\rho\,d\phi}
{\int S_0\,\rho\,d\rho\,d\phi}
\right]^{1/2}.
\label{eq:texture_contrast}
\end{equation}
By construction, $\mathcal{C}=0$ for a beam with perfectly homogeneous linear polarization
across its profile, and $\mathcal{C}>0$ whenever torsion generates a nonuniform transverse
texture. This definition, therefore, measures the formation of spatial texture rather than the local degree of polarization.

For the figures reported in the present work, we use two complementary levels of description.
Figures~\ref{fig:stokes_maps}--\ref{fig:combined_panel} are evaluated in the short-distance
regime in which diffraction is neglected and the field is approximated by the analytic local-phase solution of Eqs.~\eqref{eq:local_spinor_beam}--\eqref{eq:stokes_torsion}. These plots visualize the leading beam-level consequences of the torsion-induced phase splitting in its most transparent form. Figures~\ref{fig:diffraction_comparison}--\ref{fig:diffraction_regime_map}, by contrast, are obtained from the full numerical solution of Eq.~\eqref{eq:dimensionless_eq}, including transverse diffraction. Together, these two sets of results allow us to separate the geometric mechanism itself from the quantitative modifications introduced by paraxial propagation. Within this framework, nontrivial Stokes patterns are the primary torsional signature, whereas OAM spectra serve mainly as consistency checks showing that a purely radial torsional splitting does not automatically induce strong azimuthal mode conversion.

\section{Results and discussion}
\label{sec:results}

Before discussing the numerical outputs, it is useful to state clearly what the minimal model does and does not predict. Within the present framework, torsion acts by lifting the degeneracy between the two circular-polarization sectors through the local splitting
\begin{equation}
\Delta k_z \equiv k_z^{(-)}-k_z^{(+)}=2\Omega\rho.
\end{equation}
All nontrivial Stokes textures, polarization-angle maps, and effective birefringence profiles
discussed below are direct consequences of this torsion-induced helicity splitting. In the
reference limit $\Omega=0$, the splitting disappears, and the beam preserves its initial
homogeneous polarization apart from ordinary diffraction.

Accordingly, the numerical results should be interpreted as controlled manifestations of the
geometric mechanism isolated in the theory. They are intended to show how the local torsion law is encoded in finite-width beam observables, not to provide a fully material-specific simulation with complete microscopic realism. For that reason, the discussion proceeds in two steps. First, Figs.~\ref{fig:stokes_maps}--\ref{fig:combined_panel} present the analytic short-distance regime in which diffraction is neglected, thereby isolating the torsional phase imprint itself. Second, Figs.~\ref{fig:diffraction_comparison}--\ref{fig:diffraction_regime_map} solve the full paraxial model including diffraction in order to determine which signatures survive under realistic beam spreading and how the regime boundaries are shifted quantitatively.

\subsection{Transverse polarization textures}

We first analyze the transverse beam structure generated by propagation in the torsional medium. For an initially Gaussian beam with homogeneous linear polarization, the total intensity remains close to the scalar Gaussian envelope, whereas the polarization sector develops a nontrivial radial modulation governed by the accumulated torsional phase. This already shows that, within the minimal model, the dominant signature of uniform torsion is not strong intensity reshaping, but polarization structuring across the beam profile.

\begin{figure*}[t]
  \centering
  \includegraphics[width=\textwidth]{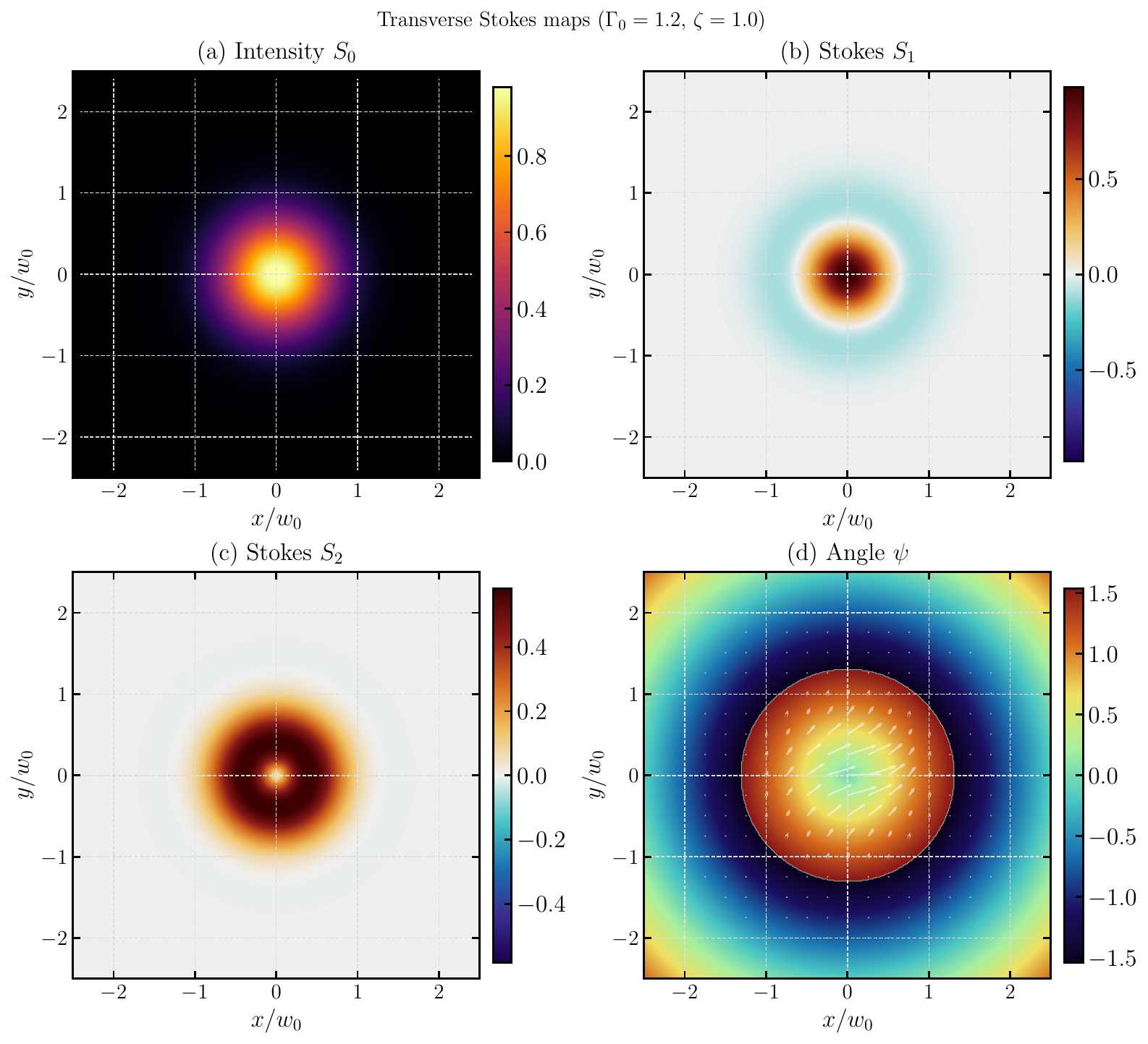}
  \caption{Transverse Stokes maps for a Gaussian input beam ($\Gamma_0=1.2$, $\zeta=1.0$):
  (a) intensity profile $S_0$,
  (b) linear-polarization Stokes parameter $S_1$,
  (c) linear-polarization Stokes parameter $S_2$, and
  (d) local polarization angle $\psi$ with polarization-direction arrows overlaid.
  The intensity remains close to the Gaussian envelope, whereas the polarization sector develops
a torsion-induced radial texture in the diffraction-neglected short-distance regime. The angle
  map in panel (d) is shown on the principal branch of $\frac12\arg(S_1+iS_2)$.}
  \label{fig:stokes_maps}
\end{figure*}

Figure~\ref{fig:stokes_maps} summarizes the four basic transverse observables for a representative Gaussian beam with $\Gamma_0=1.2$ and $\zeta=1.0$. Figure \ref{fig:stokes_maps}(a) shows the intensity distribution $S_0$, which remains essentially Gaussian and confirms that the underlying spatial beam profile is only weakly modified in the diffraction-neglected regime adopted for the figure generation. The torsion-induced response is therefore more naturally diagnosed in polarization-resolved quantities than in intensity alone.

Figures \ref{fig:stokes_maps}(b) and \ref{fig:stokes_maps}(c) display the Stokes parameters $S_1$ and $S_2$. Their annular patterns reflect the radial phase accumulation generated by the helicity splitting. In particular, $S_1 \propto \cos(2\Gamma_0 r\zeta)$ and $S_2 \propto \sin(2\Gamma_0 r\zeta)$ are shifted by $\pi/2$ relative to one another, so that together they reconstruct the local linear-polarization state across the beam cross-section. The appearance of these alternating radial domains is the most direct finite-width manifestation of the underlying radius-dependent optical activity.

Figure \ref{fig:stokes_maps}(d) presents the local polarization angle $\psi$, together with polarization-direction arrows. This panel provides the clearest geometric interpretation of the effect: the polarization axis rotates by an amount that increases with radius, in agreement with the local law $\psi=\Gamma_0 r\zeta$ modulo the usual branch convention of the complex argument. Since the angle is reconstructed numerically from $\psi=\frac12 \arg(S_1+iS_2)$, the color map is necessarily displayed on a principal branch and may therefore show apparent negative values or discontinuities. These are not additional physical structures, but simply the standard branch-cut representation of an angle variable defined modulo $\pi$. Different annular regions, therefore, accumulate different helicity-dependent phases, which produce a structured polarization texture even though the input
field is initially homogeneous. The main conclusion from Fig.~\ref{fig:stokes_maps} is that
uniform torsion acts as a spatially resolved polarization rotator, producing beam-level Stokes
textures as the robust optical consequence of the local circular birefringence.

\begin{figure*}[htbp]
    \centering
    \includegraphics[width=\textwidth]{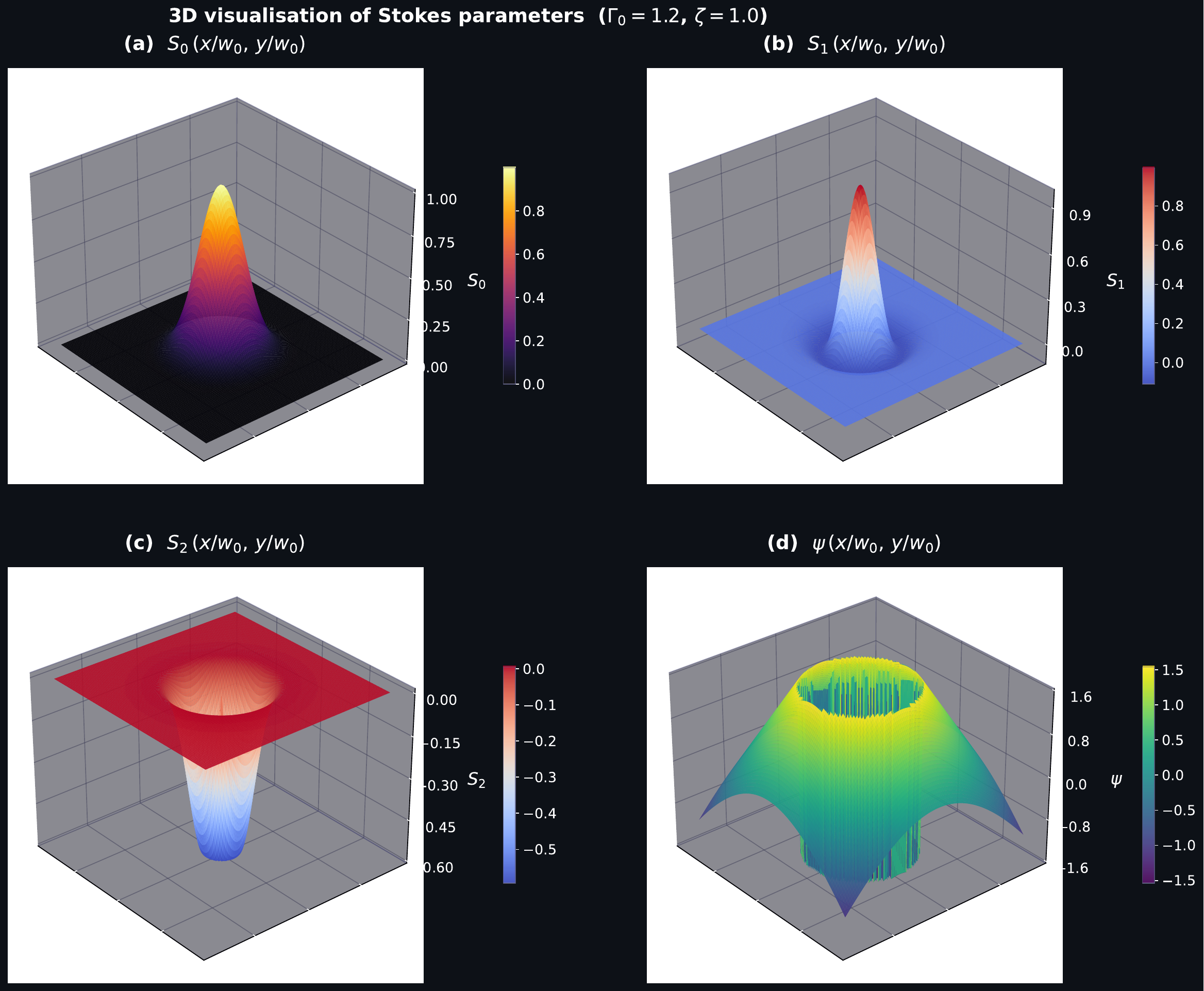}
    \caption{
        Three-dimensional visualization of the same representative beam shown in Fig.~\ref{fig:stokes_maps}, for $\Gamma_0 = 1.2$ and $\zeta = 1.0$. (a)~Total intensity $S_0(x/w_0,\,y/w_0)$, showing the preserved Gaussian envelope.
        (b)~$S_1(x/w_0,\,y/w_0)$ and (c)~$S_2(x/w_0,\,y/w_0)$, displaying the radial structure of the linear-polarization sector. (d)~Local polarization angle
        $\psi(x/w_0,\,y/w_0)=\frac12\arg(S_1+iS_2)$, emphasizing the radial growth of the torsion-induced rotation. The angular surface is shown on the principal branch and therefore exhibits the expected wrapping discontinuities.}
    \label{fig:stokes_3d}
\end{figure*}

Figure~\ref{fig:stokes_3d} provides a complementary three-dimensional visualization of the same transverse quantities shown in Fig.~\ref{fig:stokes_maps}, for the representative case
$\Gamma_0=1.2$ and $\zeta=1.0$. Its role is not to introduce a new observable, but to make the
radial morphology of the torsion-induced polarization texture more transparent.

The total intensity $S_0$ [Fig. \ref{fig:stokes_3d}(a)] retains the expected Gaussian envelope, with a dominant central peak and smooth decay toward the periphery. By contrast, the polarization-sensitive quantities $S_1$ [Fig. \ref{fig:stokes_3d}(b)] and $S_2$ [Fig. \ref{fig:stokes_3d}(c)] display nontrivial radial structure generated by the helicity-dependent phase accumulation. In particular, the three-dimensional profile of $S_2$ makes especially clear that the torsion-induced response is not a mere sign-definite modulation, but part of a phase-shifted Stokes pair $(S_1,S_2)$ encoding the local rotation of
the linear-polarization axis.

Figure \ref{fig:stokes_3d}(d) displays the local polarization angle $\psi=\frac12\arg(S_1+iS_2)$. As in the two-dimensional map, the surface is displayed on the principal branch and therefore contains the expected discontinuities associated with angular wrapping. These branch jumps are a property of the angular representation rather than an instability of the beam. Read together with Figs.~\ref{fig:stokes_maps}(b) and \ref{fig:stokes_maps}(c), panel~(d) simply confirms that the torsional phase accumulation grows with radial distance and reorganizes the transverse field into a structured polarization pattern.

The two-dimensional and three-dimensional representations, therefore, convey the same physical
message. The beam profile itself remains close to the input Gaussian envelope, but the
polarization sector becomes strongly structured through the radius-dependent relative phase between the two circular components. In this sense, the most robust effect of uniform torsion in the minimal model is the generation of beam-resolved polarization texture rather than large-scale reshaping of the scalar intensity distribution.

\subsection{Global observables and operational regimes}

\begin{figure*}[t]
  \centering
  \includegraphics[width=\textwidth]{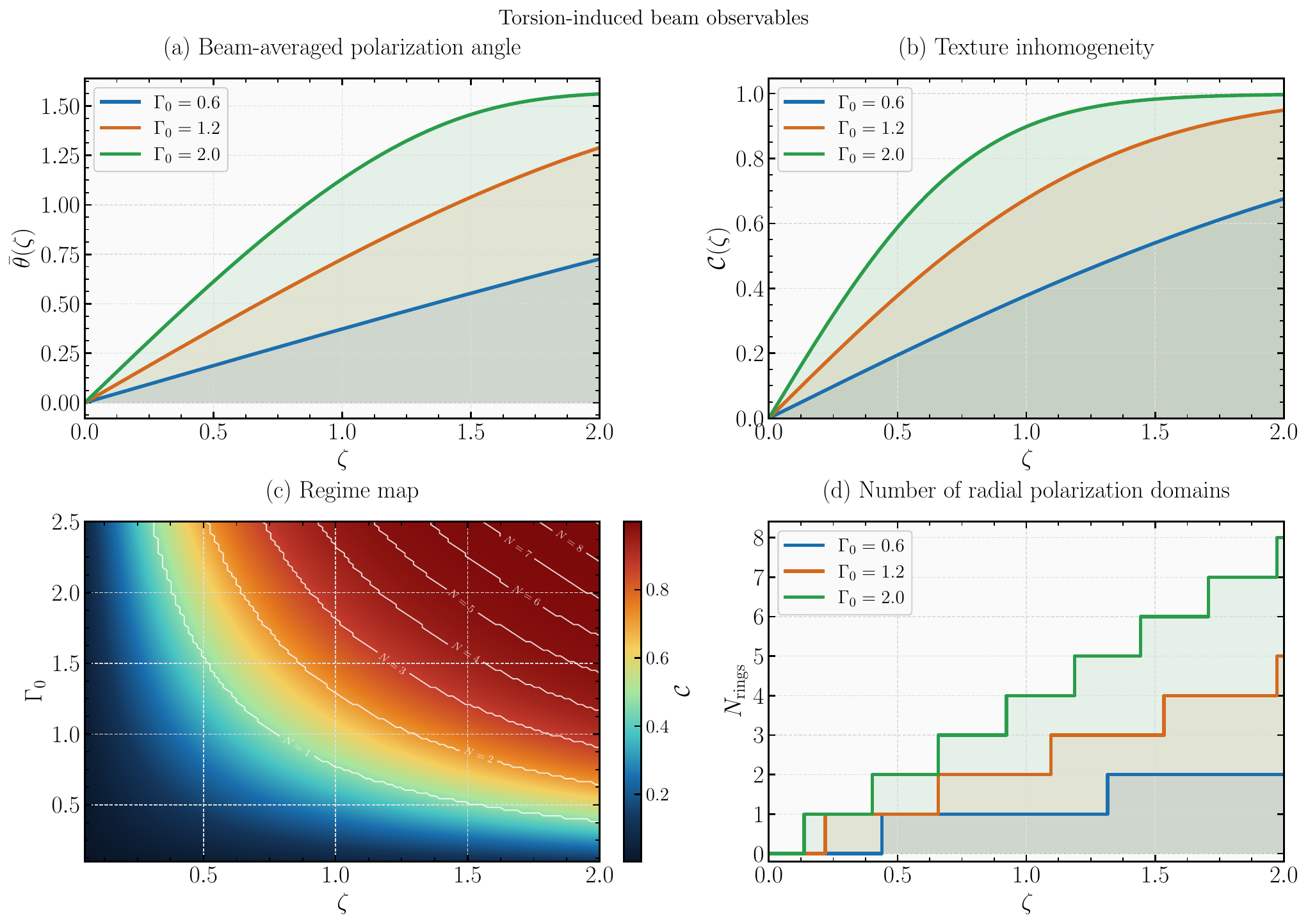}
  \caption{Global beam diagnostics in the minimal torsion model:
  (a) beam-averaged polarization angle $\bar{\theta}(\zeta)$,
  (b) texture inhomogeneity measure $\mathcal{C}(\zeta)$,
  (c) regime map $\mathcal{C}(\Gamma_0,\zeta)$ with $N_{\rm rings}$ contours, and
  (d) number of radial polarization domains $N_{\rm rings}(\zeta)$. Together these observables characterize the net output polarization angle, the visibility of the transverse texture, and the emergence of polarization domains.}
  \label{fig:combined_panel}
\end{figure*}

While the full transverse maps provide the most detailed information, it is also useful to extract global observables that summarize the beam response in compact form. Figure~\ref{fig:combined_panel} collects four complementary diagnostics: the beam-averaged
polarization angle $\bar{\theta}(\zeta)$, the texture inhomogeneity measure $\mathcal{C}(\zeta)$, the regime map in the $(\Gamma_0,\zeta)$ plane, and the number of radial polarization domains $N_{\rm rings}(\zeta)$.

These quantities are especially useful because they correspond naturally to different levels of experimental access. The angle $\bar{\theta}$ is the direct analogue of a net output polarization orientation obtained from spatially integrated polarimetry. The inhomogeneity measure $\mathcal{C}$ quantifies how far the beam has departed from a transversely uniform polarization state and is therefore naturally associated with image-based Stokes diagnostics. The domain count $N_{\rm rings}$ converts the continuously accumulated torsional phase into a discrete structural signature that can be extracted from radial sign changes in the measured linear-polarization maps. The three observables, therefore, probe complementary aspects of the same geometric mechanism.

Figure~\ref{fig:combined_panel}(a) shows the beam-averaged polarization angle $\bar{\theta}$
obtained from the integrated Stokes vector for $\Gamma_0=0.6$, $1.2$, and $2.0$. This observable measures the orientation of the \emph{net} output linear polarization. In the weak-texture regime, the curves initially track the expected geometric build-up. As the transverse texture develops, radial dephasing may produce partial cancellations in the integrated Stokes signal. In the representative curves shown here, however, the dominant feature is a smooth net rotation whose magnitude increases with $\Gamma_0$. The important point is that $\bar{\theta}$ should be interpreted as a global output angle extracted from the integrated Stokes vector, not as a strictly cumulative phase variable. The local rotation law remains linear at fixed radius, but the globally integrated signal is sensitive to how different annular regions contribute collectively to the output beam.

Figure~\ref{fig:combined_panel}(b) displays the texture inhomogeneity measure
$\mathcal{C}(\zeta)$ for the same values of $\Gamma_0$. By construction, $\mathcal{C}=0$ for a
transversely homogeneous linear-polarization pattern, which is precisely the input configuration
at $\zeta=0$. As propagation proceeds, torsion generates a nonuniform Stokes texture and
$\mathcal{C}$ becomes nonzero. Larger values of $\Gamma_0$ drive the system more rapidly into a
strongly structured regime. Depending on the parameter window, the curves may also show
nonmonotonic behavior, reflecting repeated build-up and partial washing-out of the radial pattern
under continued phase accumulation. This observable therefore measures the degree of texture
formation rather than the local degree of polarization.

Figure~\ref{fig:combined_panel}(c) synthesizes these trends in the form of a regime map in the
$(\Gamma_0,\zeta)$ plane. The color scale encodes the texture measure $\mathcal{C}$, while the
white contours separate regions with different values of the number of radial polarization domains
$N_{\rm rings}$. Three broad operating regimes emerge naturally. At small $\Gamma_0$ and short
propagation distance, the beam remains close to its initial homogeneous-polarization state,
corresponding to a weak-rotation regime. At intermediate parameter values, a small number of
well-resolved annular polarization domains appear, defining a resolved-texture regime. At larger
torsion strength or longer propagation distance, the cross-section is divided into multiple
annular domains, corresponding to a multi-domain regime. This map is best regarded as a compact
operating diagram for the minimal theory, not as a quantitative material-specific device chart.

Figure~\ref{fig:combined_panel}(d) presents the number of radial polarization domains
$N_{\rm rings}$ as a function of $\zeta$ for the same set of torsion strengths. Since the sign
inversions of $S_1 \propto \cos(2\Gamma_0 r\zeta)$ occur discretely, $N_{\rm rings}$ evolves as
a staircase function. Each step marks the appearance of an additional resolvable polarization ring
across the beam profile. This makes $N_{\rm rings}$ a particularly transparent diagnostic, because
it converts the continuously accumulated torsional phase into a directly countable beam-structure
observable.

Taken together, the four panels of Fig.~\ref{fig:combined_panel} provide a compact summary of the
minimal model. The transverse torsion-induced phase splitting generates structured Stokes patterns
locally, while globally it produces a net output polarization angle, a measurable degree of
texture inhomogeneity, a discrete domain count, and a clear separation between weak, resolved, and
multi-domain regimes. These observables, therefore, establish a practical bridge between the local
geometric mechanism and beam-level diagnostics.

% ------------------------------------------------------------------
\subsection{Scaling and physical interpretation}
% ------------------------------------------------------------------

The numerical results make explicit a clear hierarchy of effects. At the most local level, the
torsion-induced splitting generates a radius-dependent phase difference between the two circular
sectors, which is directly encoded in the radial oscillation of $S_1$ and $S_2$ and in the growth
of the local angle $\psi$. At the beam level, this local phase structure gives rise to a nonzero
texture inhomogeneity measure $\mathcal{C}$ and to the emergence of discrete annular polarization
domains. Finally, when the field is reduced to an integrated Stokes signal, the net output
polarization is summarized by $\bar{\theta}$, which is sensitive not only to the local geometric
rotation but also to cancellations produced by radial dephasing.

This hierarchy is useful because it separates three physically distinct levels of description:
local geometric birefringence, resolved transverse texture formation, and net integrated
polarimetric response. In particular, a small value of $\bar{\theta}$ does not necessarily imply a
weak local torsion effect; it may also reflect partial cancellation between different annular
regions of a strongly structured beam. Conversely, a nonzero $\mathcal{C}$ directly certifies
that the output field has acquired a spatially inhomogeneous polarization texture even when the
integrated Stokes vector is partially suppressed. The pair $(\bar{\theta},\mathcal{C})$ is
therefore operationally richer than either quantity taken in isolation.

An important consequence inherited from the local theory is the achromatic character of the
geometric contribution within the ideal model. Since $\Delta n = 2c\Omega\rho/\omega$ and the standard optical-activity relation gives $\Delta\theta=(\pi L/\lambda)\Delta n$, the explicit wavelength dependence cancels, yielding the geometric law $\Delta\theta=\Omega\rho L$. Thus, in the idealized framework considered here, the torsion-induced rotation is achromatic. Any additional material dispersion, absorption, or non-geometric birefringence would lie beyond the scope of the present minimal model.

The results also clarify the limitations of the cylindrically symmetric setting. The dominant and most robust predictions are geometric birefringence, differential polarization rotation, Stokes-texture formation, and the emergence of radial polarization domains. By contrast, strong orbital-angular-momentum conversion is not expected as a generic outcome, because the torsion-induced phase in the minimal model depends only on radius and does not introduce a new azimuthal winding. In this sense, the present framework identifies polarization structuring as the primary structured-light signature of uniform torsion, whereas OAM redistribution remains a higher-level possibility requiring additional geometric or photonic structure.

\subsection{Including diffraction: robustness of the polarization textures}
\label{sec:diffraction}

The results discussed in Figs.~\ref{fig:stokes_maps}--\ref{fig:combined_panel} were obtained in the short-distance, diffraction-neglected regime, which isolates the geometric phase mechanism by neglecting the transverse Laplacian. This approximation is appropriate for propagation distances
$z \ll z_R = k_0 w_0^2$, where the beam profile evolves only weakly. Figures~\ref{fig:diffraction_comparison}--\ref{fig:diffraction_regime_map}, however, are obtained from the full paraxial model, in which diffraction and torsion-induced helicity splitting act simultaneously. In this subsection, we solve
\begin{equation}
i\partial_\zeta \Psi =
\left[
-\frac{1}{2}\nabla_r^2 I + \Gamma_0 r \,\sigma_3
\right]\Psi,
\label{eq:dimensionless_full}
\end{equation}
where $\nabla_r^2=\partial_r^2+r^{-1}\partial_r+r^{-2}\partial_\phi^2$, and $\Gamma_0$ is the
dimensionless torsion parameter defined in Eq.~\eqref{eq:Gamma_linear_torsion}. Our goal is to
test the robustness of the torsion-induced polarization textures beyond the analytic short-distance
limit and to determine how diffraction modifies the sharpness of the annular domains and the
location of the operating-regime boundaries.

\subsubsection{Numerical methodology}

Equation~\eqref{eq:dimensionless_full} is solved numerically on a radial grid adapted to the cylindrically symmetric case relevant for the present figures. For the Gaussian input states considered here, only the azimuthally symmetric sector $m=0$ is populated initially, and the torsion term $\Gamma_0 r\sigma_3$ remains diagonal in the circular basis. The problem, therefore, reduces to a coupled set of radial evolution equations for the two circular envelopes.

We discretize the radial Laplacian on a uniform grid $r_j=j\,\Delta r$ with $j=0,\dots,N_r-1$, using the standard finite-difference form appropriate to cylindrical symmetry and imposing regularity at the origin together with a sufficiently large outer cutoff $r_{\max}$ to suppress spurious boundary effects. Diffraction is then propagated with a Crank--Nicolson step, while the torsion-induced helicity splitting is applied through a symmetric phase update in the circular basis. In practice, one evolution step is implemented as a symmetric split operation: half a torsional phase step, one Crank--Nicolson diffraction step, and a final half torsional phase step. This scheme preserves the helicity-diagonal structure of the torsion term and provides stable paraxial propagation over the full $\zeta$ range used in Figs.~\ref{fig:diffraction_comparison}--\ref{fig:diffraction_regime_map}.

The input beam is taken to be a Gaussian radial envelope normalized in total power rather than in peak amplitude, so that the radial integral of $S_0$ is fixed. For this reason, the plotted peak values of $S_0$, $S_1$, and $S_2$ in the radial profiles need not be bounded by unity. The diffraction-neglected comparison curves are generated analytically from the same normalized input envelope,
\begin{align}
S_1(r)=|u_0(r)|^2\cos(2\Gamma_0 r\zeta),\\
S_2(r)=|u_0(r)|^2\sin(2\Gamma_0 r\zeta),
\end{align}
so that all comparisons are conducted using a consistent normalization convention.

\subsubsection{Comparison with the diffraction-neglected regime}

\begin{figure*}[t]
  \centering
  \includegraphics[width=\textwidth]{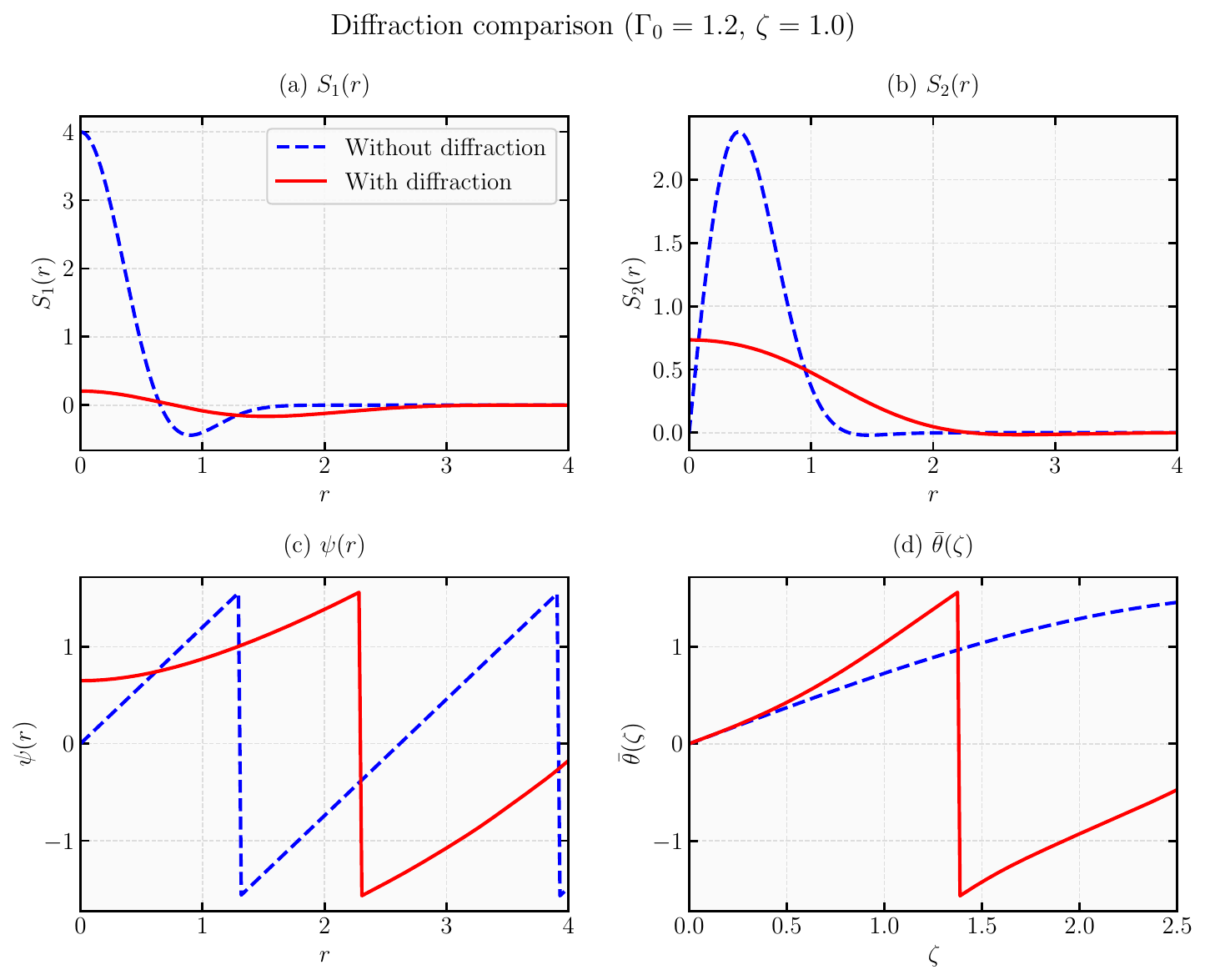}
  \caption{
    Comparison between the diffraction-neglected regime (dashed lines) and the full paraxial
    propagation including diffraction (solid lines) for $\Gamma_0 = 1.2$ at $\zeta = 1.0$.
    (a) Radial profile of $S_1(r)$, (b) radial profile of $S_2(r)$, (c) local polarization
    angle $\psi(r)$, and (d) beam-averaged polarization angle $\bar{\theta}(\zeta)$.
    Both cases use the same total-power-normalized Gaussian input envelope, so the plotted peak
    values are not constrained to unity. Diffraction smooths the sharp transitions between
    polarization domains but preserves the overall radial structure. The local rotation rate
    remains close to $\Gamma_0 \zeta$, and the global angle $\bar{\theta}(\zeta)$ follows
    the diffraction-neglected curve closely over the initial propagation range.}
  \label{fig:diffraction_comparison}
\end{figure*}

Figure~\ref{fig:diffraction_comparison} compares the radial profiles of the Stokes parameters and
the local polarization angle obtained from the full paraxial equation (solid lines) with those
from the diffraction-neglected approximation (dashed lines) for $\Gamma_0 = 1.2$ at $\zeta = 1.0$.
The dashed comparison curves are generated analytically from the same total-power-normalized input
envelope used in the full simulation. Consequently, the peak values of the displayed radial Stokes
profiles are not constrained to unity.

Figure~\ref{fig:diffraction_comparison}(a) shows $S_1(r)$ and Fig.~\ref{fig:diffraction_comparison}(b) shows $S_2(r)$. The full solution preserves the overall oscillatory structure but exhibits smoother transitions between positive and negative lobes. This smoothing is a direct consequence of the transverse coupling introduced by the Laplacian: neighboring radial points exchange amplitude via diffraction, blurring the sharp phase boundaries that would otherwise exist. In the representative case shown here, the main sequence of sign changes is preserved, indicating that the radial-domain structure remains identifiable even after diffractive smoothing.

Figure~\ref{fig:diffraction_comparison}(c) displays the local polarization angle $\psi(r) = \frac{1}{2}\arg(S_1+iS_2)$. In the diffraction-neglected case, $\psi(r)$ follows the linear law $\psi(r) = \Gamma_0 r \zeta$ exactly. With diffraction included, $\psi(r)$ still increases approximately linearly with $r$, but small deviations appear near the beam center and at the boundaries between domains, where the smoothing effect is most pronounced. Importantly, the average slope remains close to $\Gamma_0 \zeta$, confirming that the local rotation rate is preserved to good approximation.

Figure~\ref{fig:diffraction_comparison}(d) presents the beam-averaged polarization angle $\bar{\theta}(\zeta)$ for the two cases. The full solution (solid) tracks the diffraction-neglected curve (dashed) closely up to $\zeta \approx 1.0$, after which a slight divergence appears. This divergence arises because diffraction redistributes intensity radially, modifying the weighting of different annular regions in the integrated Stokes vector. However, the qualitative behavior, monotonic increase with $\zeta$ followed by partial saturation, is unchanged.

These results demonstrate that the diffraction-neglected approximation captures the essential
physics of the torsion-induced polarization structuring, and that the inclusion of diffraction
introduces mainly quantitative modifications. The robustness of the textures against transverse
spreading is encouraging from the experimental point of view, where some degree of beam broadening
is unavoidable.

\subsubsection{Parameter dependence and regime shifts}

\begin{figure*}[t]
  \centering
  \includegraphics[width=\textwidth]{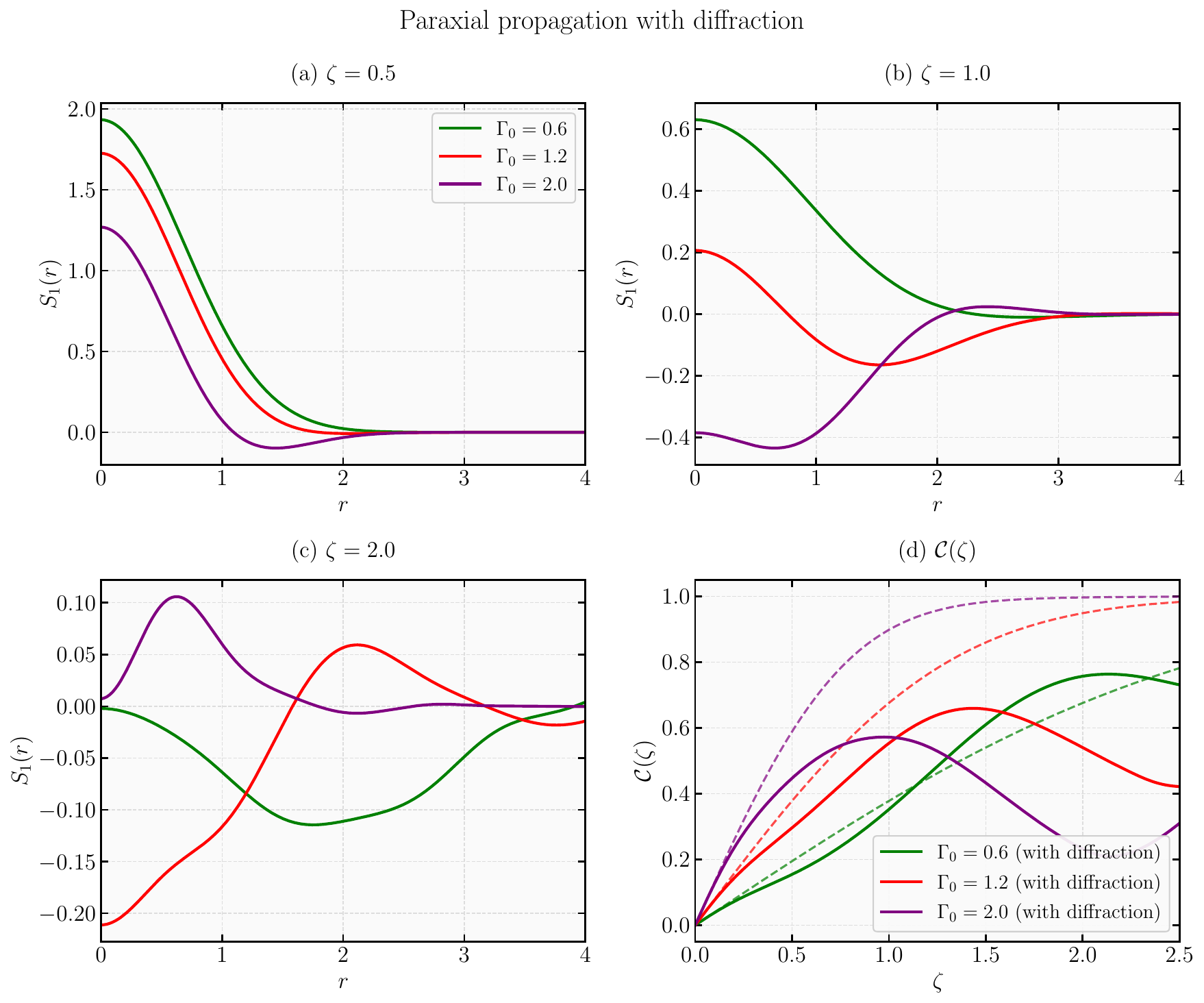}
  \caption{
    Full paraxial propagation including diffraction for $\Gamma_0 = 0.6$, $1.2$, and $2.0$.
    (a)--(c) Radial profiles of $S_1(r)$ at $\zeta = 0.5$, $1.0$, and $2.0$.
    (d) Texture inhomogeneity measure $\mathcal{C}(\zeta)$ comparing the full solution
    (solid) with the diffraction-neglected approximation (dashed). Diffraction reduces
    the peak inhomogeneity but does not eliminate the texturing effect. The qualitative
    dependence on $\Gamma_0$ and $\zeta$ remains unchanged.}
  \label{fig:diffraction_evolution}
\end{figure*}

Having established that diffraction does not destroy the polarization textures, we now investigate how the observable quantities depend on the dimensionless torsion strength $\Gamma_0$ and propagation distance $\zeta$ in the full paraxial model. Figure~\ref{fig:diffraction_evolution} explores this parameter space systematically.

Figures~\ref{fig:diffraction_evolution}(a)--\ref{fig:diffraction_evolution}(c) show the radial profiles of $S_1(r)$ at $\zeta = 0.5$, $1.0$, and $2.0$ for three representative torsion strengths: $\Gamma_0 = 0.6$, $1.2$, and $2.0$. For $\Gamma_0 = 0.6$ (weak torsion), diffraction dominates, and the oscillatory structure is only weakly visible; $S_1(r)$ remains predominantly positive, indicating that the polarization has not developed fully resolved annular domains. For $\Gamma_0 = 1.2$ (intermediate torsion), clear oscillations appear by $\zeta = 1.0$, and by $\zeta = 2.0$ multiple rings are discernible. For $\Gamma_0 = 2.0$ (strong torsion), the oscillations are well developed even at $\zeta = 0.5$, and many rings are present at longer distances. The qualitative progression is the same as in the diffraction-neglected regime, but the sharpness of the oscillations is reduced, and the visibility of higher-order rings is diminished by transverse smoothing.

Figure~\ref{fig:diffraction_evolution}(d) quantifies this behavior through the texture inhomogeneity measure $\mathcal{C}(\zeta)$ defined in Eq.~\eqref{eq:texture_contrast}. Solid lines show the full solution for the three values of $\Gamma_0$, while dashed lines show the diffraction-neglected approximation. For all $\Gamma_0$, $\mathcal{C}(\zeta)$ initially grows, reaches a maximum, and then slowly decays. The peak value is reduced by diffraction, more so for smaller $\Gamma_0$, where the diffusive scale is shorter relative to the torsional phase accumulation. However, the position of the maximum and the overall shape of the curves remain similar to the diffraction-neglected case, confirming that the measure $\mathcal{C}$ remains a useful diagnostic even when diffraction is included.

\subsubsection{Updated regime map}

\begin{figure}[t]
\centering
\includegraphics[width=0.47\textwidth]{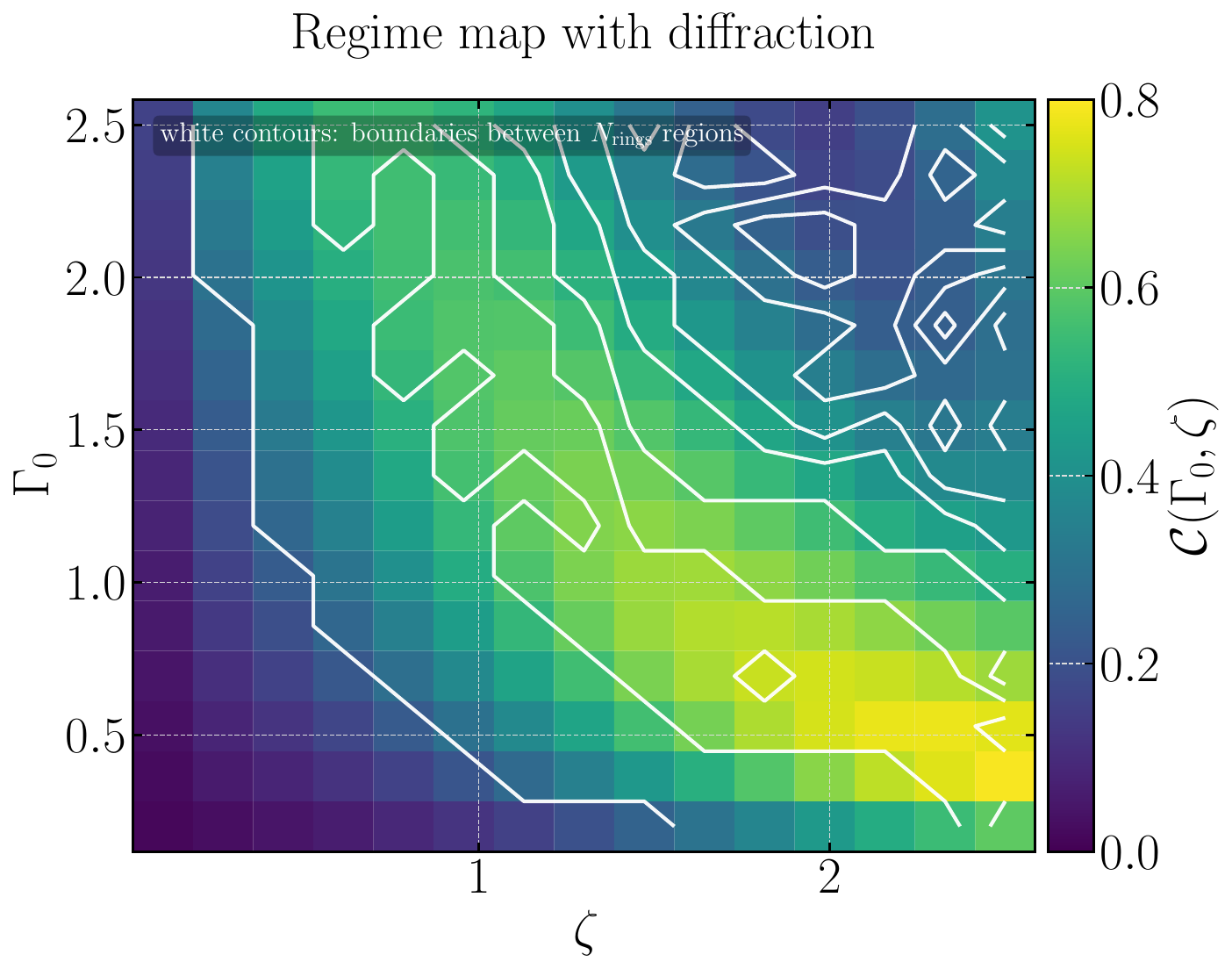}
\caption{Regime map in the $(\Gamma_0,\zeta)$ plane including diffraction. The color scale
encodes the texture measure $\mathcal{C}$ obtained from the full paraxial equation. The white
contours mark the boundaries between regions with different values of the number of radial
polarization domains $N_{\rm rings}$, as extracted numerically from the zero-crossing structure of
$S_1(r)$. Compared with the diffraction-neglected map of Fig.~\ref{fig:combined_panel}(c), the
regime boundaries are shifted toward larger $\Gamma_0$ or $\zeta$ by a factor of order
$1.5$--$2$, reflecting the smoothing effect of transverse coupling. The qualitative structure
remains unchanged. The figure should be read as an operational threshold map: the color scale shows the strength of texture formation through $C(\Gamma_0,\zeta)$, while the white contours indicate the onset of additional resolved radial domains.}
\label{fig:diffraction_regime_map}
\end{figure}

The regime map presented in Fig.~\ref{fig:combined_panel}(c) was constructed in the
 diffraction-neglected approximation and therefore overestimates the sharpness of the boundaries
between the weak, resolved, and multi-domain regimes. Figure~\ref{fig:diffraction_regime_map}
provides an updated map obtained from the full paraxial equation, where the color scale encodes
the texture measure $\mathcal{C}(\Gamma_0,\zeta)$ and the white contours mark the boundaries
between regions with different values of the radial-domain count $N_{\text{rings}}$, extracted
operationally from the zero-crossing structure of $S_1(r)$ within the resolved-intensity support
of the beam.

Compared with the earlier map, the boundaries are shifted toward larger $\Gamma_0$ or $\zeta$ by
a factor of approximately $1.5$--$2$. This shift reflects the additional length scale introduced
by diffraction: the beam must propagate further to accumulate enough torsional phase to overcome
the smoothing effect of transverse coupling. In other words, diffraction does not destroy the
structured-light signature of torsion, but it raises the threshold for a clearly resolved texture.

Despite these quantitative shifts, the qualitative structure of the map remains intact. Three distinct operating regimes --- weak, resolved, and multi-domain --- are still identifiable, and the full-diffraction map continues to organize the parameter space in a way consistent with the basic phase-accumulation scaling. The updated map, therefore, serves as a more realistic guide for experimental design, accounting for the inevitable transverse spreading of real paraxial beams.

\subsubsection{Discussion and implications}

The numerical results presented in this subsection lead to several important conclusions. First, the polarization textures predicted by the minimal torsion model are robust against diffractive effects. While diffraction smooths the sharp transitions between annular domains and reduces the peak value of the inhomogeneity measure $\mathcal{C}$, it does not destroy the underlying radial structure. The local polarization angle $\psi(r)$ continues to increase approximately linearly with $r$, and the radial-domain structure remains well defined over the parameter ranges relevant to Figs.~\ref{fig:diffraction_comparison}--\ref{fig:diffraction_regime_map}.

Second, the global observables $\bar{\theta}(\zeta)$ and $\mathcal{C}(\zeta)$ retain their operational meaning even in the presence of diffraction. The beam-averaged angle tracks the diffraction-neglected curve closely over the initial propagation window, after which deviations arise from radial intensity redistribution. The inhomogeneity measure provides a quantitative metric for texture formation that is sensitive to both torsion strength and propagation distance, and its qualitative behavior, initial growth, saturation, and slow decay, persists with diffraction.

Third, the updated regime map of Fig.~\ref{fig:diffraction_regime_map} offers a more realistic guide for experimental design. The shift of the regime boundaries toward larger $\Gamma_0$ or $\zeta$ implies that achieving a given number of polarization domains requires either stronger torsion (larger $\Omega$), larger beam waists (increasing $z_R$ and hence $\Gamma_0$ for fixed $\Omega$), or longer propagation distances. This trade-off is captured quantitatively by the map and can be used to optimize experimental parameters. Operationally, Fig.~\ref{fig:diffraction_regime_map} should be read as a threshold map: the color scale measures how strongly the polarization texture has developed through $C(\Gamma_0,\zeta)$, while the white contours indicate when additional resolved radial domains appear. Thus, increasing either $\Gamma_0$ or $\zeta$ drives the beam toward more visible and more highly structured textures, although diffraction shifts these thresholds upward relative to the diffraction-neglected map.

Finally, the robustness of the polarization textures against diffraction strengthens the case for experimental realization of the torsion-induced effects discussed in Sec.~\ref{sec:applications}. The fact that the key qualitative features survive transverse coupling confirms that they are not artifacts of the diffraction-neglected approximation, but genuine physical consequences of the helicity splitting $\Delta k_z = 2\Omega\rho$. Any experimental platform that realizes the effective Hamiltonian $\Omega\rho\sigma_3$ should exhibit the polarization structuring described here, subject to the quantitative modifications identified in this section.

In summary, the inclusion of diffraction confirms and refines the predictions of the minimal model, demonstrating that the torsion-induced polarization textures are observable in realistic paraxial beams. The numerical results bridge the gap between idealized geometric theory and practical experimental implementation, and establish a clear hierarchy of effects: local helicity splitting $\to$ transverse texture formation $\to$ global polarimetric signatures, all of which remain meaningful when diffraction is taken into account.

\section{Discriminating signatures and routes beyond the minimal radial model}
\label{sec:beyond}

The results obtained so far establish a controlled baseline for the optical consequences of uniform torsion. They also make it possible to separate two different questions that are often merged in structured-light discussions. The first is whether torsion generates a measurable beam-resolved optical signature at all. The present answer is clearly yes: the local splitting $\Delta k_z=2\Omega\rho$ produces radial polarization textures, nonzero $\mathcal{C}$, and a well-defined sequence of polarization domains. The second question is whether such a signature is already sufficient to identify a genuinely geometric torsional mechanism, or whether similar textures could arise from more conventional forms of spatially varying optical activity. This section addresses that distinction and, in doing so, clarifies what extra structure would be needed to move from pure radial texturing to genuine torsion-assisted spin--orbit conversion.

\subsection{Discriminating the torsional signature from conventional gyrotropy}

A homogeneous optically active medium provides a useful reference case. If the circular birefringence were spatially uniform, the two helicity sectors would accumulate only a constant relative phase across the beam profile. An initially homogeneous Gaussian beam would then remain transversely uniform in polarization, apart from ordinary diffraction, so that the texture diagnostic $\mathcal{C}$ would ideally remain zero and no radial polarization domains would be generated. In this sense, any nonzero $\mathcal{C}$ or nontrivial $N_{\rm rings}$ produced from a homogeneously polarized input beam already signals that the helicity splitting varies across the transverse plane.

At the same time, a more general \emph{non-geometric} medium with a deliberately engineered radial gyrotropic gradient could reproduce beam-level patterns similar to those of the minimal torsion model. For that reason, the present observables should not be interpreted as absolute proof of Riemann--Cartan torsion on their own. What the geometric model contributes is a specific and internally consistent scaling hierarchy in which the same torsional parameter controls the local splitting, the local polarization angle, the dimensionless beam parameter, and the global diagnostics:
\begin{equation}
\Delta k_z = 2\Omega\rho,
\qquad
\psi(\rho,z)=\Omega\rho z,
\qquad
\Gamma_0 = z_R \Omega w_0.
\label{eq:torsion_scaling_hierarchy}
\end{equation}
A meaningful experimental test of the geometric model should therefore check not only the presence
of radial polarization rings, but the joint scaling of $\bar{\theta}$, $\mathcal{C}$, and
$N_{\rm rings}$ with propagation distance, beam waist, and representative beam radius.

\subsection{What is missing for genuine spin--orbit conversion?}

The present calculations also make explicit why pronounced orbital-angular-momentum conversion is
absent in the minimal cylindrically symmetric model. The torsion-induced phase splitting enters as
a purely radial helicity-diagonal term, proportional to $\Omega\rho\,\sigma_3$, so it rotates the
polarization locally without imposing a new azimuthal winding. To go beyond this regime, the beam
must experience an effective geometric coupling with nontrivial azimuthal structure.

A useful way to state this is to consider a local polarization-frame transformation
\begin{equation}
\Psi(\rho,\phi,\zeta)=U(\rho,\phi)\,\widetilde{\Psi}(\rho,\phi,\zeta),
\end{equation}
under which the paraxial beam equation acquires the geometric gauge field
\begin{equation}
\mathbf{A}_g = i\,U^{\dagger}\nabla_{\perp}U.
\label{eq:geometric_connection_newsec}
\end{equation}
At the effective level, the transformed beam then obeys a generalized propagation law of the form
\begin{equation}
i\partial_{\zeta}\widetilde{\Psi}=
\left[
-\frac{1}{2}\left(\nabla_{\perp}-i\mathbf{A}_g\right)^2
+\widetilde{V}(\rho,\phi)
\right]\widetilde{\Psi},
\label{eq:generalized_beam_connection}
\end{equation}
where $\widetilde{V}$ denotes the transformed helicity-dependent potential together with possible
additional scalar terms. In this language, strong azimuthal mode conversion requires more than the
radial splitting itself; it requires either an explicitly $\phi$-dependent effective potential or a
nonvanishing azimuthal component $A_{g,\phi}$ of the geometric connection.

This criterion gives a concrete roadmap for future extensions of the theory. Three routes are particularly natural: (i) a torsion profile $\Omega(\rho,\phi)$ that breaks cylindrical symmetry,
(ii) an effective constitutive realization whose local optical axes or gyrotropic tensor depend on
azimuth as well as radius, or (iii) a more complete beam theory in which the transport of the
local polarization basis generates a nontrivial geometric connection. In all three cases, the
expected novelty would not be merely stronger radial texturing, but the appearance of true
spin--orbit signatures such as OAM sidebands, azimuthally structured Stokes textures, or
torsion-assisted geometric phase windings.

The present work does not solve those generalized models numerically, and it would be misleading to
claim otherwise. What it does provide is the baseline needed to interpret them correctly. By
showing that uniform torsion already produces robust beam-resolved polarization structuring but
not strong OAM conversion, the paper identifies the missing structural ingredient --- an effective
azimuthal geometric connection --- that must be added before a higher-level torsion-assisted
spin--orbit phenomenology can emerge. This makes the current manuscript stronger conceptually even
within its controlled minimal scope, because it delineates both what has been established and what
would constitute a genuine next-step advance.

\section{Operational implications and experimental outlook}
\label{sec:applications}

The effective torsional model developed here suggests a simple qualitative rule for interpreting
beam propagation within the regime of validity of the minimal description. Since the local
polarization rotation obeys
\begin{equation}
\Delta\theta=\Omega\rho L,
\end{equation}
as given already in Eq.~\eqref{eq:rotation_law}, a target geometric rotation at a representative
beam radius fixes the required value of $\Omega\rho$. Through $\Omega=b\sigma/2$, this may be
translated into a constraint on the Burgers-vector density of the underlying defect distribution,
or, more generally, into an effective torsional parameter for an engineered photonic platform
designed to emulate the spiral geometry.

\subsection{Proposed macroscopic realization}

While the torsion parameter $\Omega$ was originally introduced through the geometric model of screw
dislocations, its physical essence, a radially dependent circular birefringence,
$\Delta n(\rho)=2c\Omega\rho/\omega$, can be realized in purely classical, macroscopic optical
media without invoking atomic-scale defects.

Consider a medium with a helicoidal structure whose optical activity varies linearly with the
radial coordinate. Such a medium can be described by a dielectric tensor of the form
\begin{equation}
\varepsilon_{ij}(\rho) = \varepsilon_0 \begin{pmatrix}
n_0^2 & i\alpha\rho & 0 \\
-i\alpha\rho & n_0^2 & 0 \\
0 & 0 & n_e^2
\end{pmatrix},
\label{eq:dielectric_tensor}
\end{equation}
where $\alpha$ is a constant characterizing the radial gradient of the gyrotropic response. This
tensor yields a local circular birefringence $\Delta n(\rho) = \alpha\rho/n_0$, which matches the
torsion-induced expression, provided we identify
\begin{equation}
\Omega = \frac{\alpha\omega}{2c n_0}.
\label{eq:omega_alpha}
\end{equation}
In this sense, the torsional parameter can be reinterpreted operationally as a compact encoding of
a radial gradient of circular birefringence.

Several classes of macroscopic platforms can in principle, approximate such a response:
femtosecond-laser-written transparent media with spatially programmed birefringent structure,
liquid-crystal cells with annular electrode control, and composite photonic or metamaterial
samples engineered to exhibit a radial gyrotropic gradient. The present theory does not require
that these platforms reproduce the full defect geometry microscopically. What matters at the beam
level is the effective helicity-diagonal term proportional to $\Omega\rho\sigma_3$.

For visible light and a moderate refractive index, the values used in the illustrative figures
correspond to a weak but experimentally accessible radial gradient. The relevant point is not the
exact numerical estimate, which remains platform dependent, but the scaling: increasing the
observable effect may be achieved by increasing $\Omega$, by increasing the beam waist $w_0$ and
hence the Rayleigh range $z_R$, or by increasing the propagation distance $L$ while remaining
within the validity window of the paraxial model.

The diffraction-inclusive regime map of Fig.~\ref{fig:diffraction_regime_map} is especially useful
in this context because it translates that scaling into an operational design chart. In the full
paraxial problem, obtaining a given number of resolved radial polarization domains requires
somewhat larger values of $\Gamma_0$ or $\zeta$ than in the diffraction-neglected limit. This
shift should be interpreted not as a failure of the geometric mechanism, but as the quantitative
cost of transverse smoothing in realistic beams.

\subsection{Experimental observables and measurement strategies}

From an experimental perspective, the most natural observables are precisely the three beam-level
quantities introduced above. The beam-averaged angle $\bar{\theta}$ is the direct analogue of a
net output polarization orientation extracted from integrated polarimetry. The inhomogeneity
measure $\mathcal{C}$ captures how strongly the output beam departs from a transversely uniform
polarization state and is therefore naturally associated with spatially resolved Stokes imaging.
The discrete count $N_{\rm rings}$ converts the continuous torsion-induced phase accumulation
into a directly interpretable morphological signature in the radial polarization texture.
Together, these three quantities provide complementary access to weak-rotation, resolved-texture,
and multi-domain regimes.

This separation of observables is practically important. In the weak-texture regime,
$\bar{\theta}$ provides the most immediate global readout of the geometric response. Once the
output beam enters a more strongly structured regime, however, the pair
$(\bar{\theta},\mathcal{C})$ becomes more informative than either quantity alone, because a
small integrated angle may coexist with a pronounced local texture owing to radial cancellations.
The domain count $N_{\rm rings}$ then supplies an additional discrete indicator of how far the
beam has progressed into the structured regime.

A realistic measurement protocol would therefore proceed in stages. A first diagnostic layer would
use integrated polarimetry to estimate $\bar{\theta}$ and to identify the weak-rotation regime. A
second layer would use spatially resolved Stokes imaging to reconstruct $S_1$ and $S_2$ across
the output beam, from which both $\mathcal{C}$ and the domain structure can be extracted. In this
setting, the diffraction-inclusive regime map becomes operationally valuable because it indicates
which combinations of torsion strength and propagation distance are expected to yield merely a net
rotation, a clearly resolved annular texture, or a multi-domain output pattern.

The same formalism also clarifies the beam-optics interpretation of the geometric action. At fixed
radius, the polarization evolution is governed by the unitary
\begin{equation}
U(\Omega,\rho,L)=e^{-i\Omega\rho L\sigma_3},
\end{equation}
which acts as a geometric phase operation in polarization space. In the ideal model, this action
is broadband at the level of the minimal geometric description, since the explicit wavelength
dependence cancels in $\Delta\theta=\Omega\rho L$. This observation connects the present geometric framework with passive polarization control, while keeping explicit that a realistic implementation would require platform-specific modeling beyond the present scope.

A realistic quantitative comparison with experiment would therefore require several extensions:
material dispersion, absorption, incomplete cylindrical symmetry, non-geometric birefringence, and
finite detector resolution. None of these ingredients changes the main role of the present
section, which is more modest and more robust, to show that the geometric torsion parameter can be
mapped onto experimentally interpretable beam observables within a controlled minimal theory.

\section{Conclusions}
\label{sec:conclusions}

We have developed a minimal effective finite-width beam description anchored in the same
torsion-induced helicity splitting that arises from Maxwell electrodynamics in a Riemann--Cartan
background with uniform torsion. In the underlying local analysis, the contortion term lifts the
degeneracy between the two circular-polarization sectors, producing the splitting
$\Delta k_z \equiv k_z^{(-)}-k_z^{(+)}=2\Omega\rho$ and the corresponding geometric rotation law
$\Delta\theta=\Omega\rho L$. The present work extends that local mechanism to a controlled
paraxial setting in which the torsion parameter remains explicit throughout the beam dynamics.

Within this cylindrically symmetric minimal model, the main physical consequence of uniform
torsion is the formation of spatially varying polarization textures across the transverse beam
profile. The resulting maps of $S_1$, $S_2$, and $\psi$ provide the beam-resolved manifestation of
the same geometric optical activity obtained in the local plane-wave picture. The associated
effective birefringence $\Delta n=2c\Omega\rho/\omega$ further clarifies the geometric origin of
the local circular splitting and the achromatic character of the ideal torsion-induced rotation
law.

A central outcome of the beam-level formulation is the identification of three complementary
observables with direct operational meaning: the beam-averaged polarization angle extracted from
the integrated Stokes vector, the texture inhomogeneity measure, and the number of radial
polarization domains. Taken together, these quantities connect the underlying torsion parameter
with experimentally interpretable output-beam signatures and make explicit the distinction between
net polarimetric rotation and resolved transverse texture formation.

The paper combines two complementary levels of description. In the analytic short-distance, diffraction-neglected regime, the geometric mechanism can be isolated in closed form, and its beam-level consequences can be read directly from the Stokes parameters. In the full paraxial model, including diffraction, the same torsion-induced textures remain robust, although the annular domains are smoothed and the regime boundaries are quantitatively shifted. This two-level structure is important because it shows that the predicted signatures are not artifacts of the simplified analytic limit.

Our analysis also shows that strong orbital-angular-momentum conversion is not a generic
prediction of the minimal uniform-torsion model. Because the torsion-induced phase splitting is
purely radial in the present framework, the most robust signature is polarization structuring
rather than pronounced azimuthal mode redistribution. In this sense, the OAM analysis plays mainly
a consistency role here, whereas substantial spin--orbit conversion should be regarded as a
possible extension requiring additional azimuthal geometric structure or more general effective
gauge couplings.

More specifically, the analysis identifies the structural ingredient needed to go beyond pure
radial texturing: an effective azimuthal geometric connection, or equivalently, an explicit
$\phi$-dependence in the helicity-dependent transport. This observation clarifies why strong OAM
sidebands are absent in the present model and provide a concrete roadmap for future extensions in
which nonuniform torsion, broken cylindrical symmetry, or polarization-basis transport could
generate genuine torsion-assisted spin--orbit conversion.

Two caveats delimit the scope of the present results. First, the beam equation is an effective
paraxial reduction anchored in the local helicity splitting, not a full exact solution of the
complete Riemann--Cartan Maxwell problem for arbitrary finite beams. Second, although the present
manuscript already includes full paraxial propagation with diffraction, it is still not a
quantitative material-specific beam-propagation theory with complete constitutive realism. These
caveats do not weaken the main conclusion of the work; rather, they clarify exactly which claims
follow robustly from the minimal theory.

The present framework, therefore, provides a physically transparent route from local torsion-induced
optical activity to finite-width beam observables. More broadly, it identifies a clear hierarchy
of effects associated with uniform torsion: local circular birefringence, radius-dependent
polarization rotation, beam-resolved Stokes textures, compact global diagnostics, and robust
paraxial signatures that survive diffraction. This makes the theory a useful starting point for
future studies incorporating realistic constitutive response, non-ideal defect distributions,
absorption, and torsion-assisted spin--orbit couplings in photonic or metamaterial platforms.

\section*{Acknowledgments}

This work was supported by CAPES (Finance Code 001), CNPq (Grant PQ/306308/2022-3), and FAPEMA (Grants UNIVERSAL-06395/22).

\appendix

\section{Derivation of the representative input states and of the short-distance factorization}
\label{app:eq33_eq34}

For completeness, we collect here the short calculations underlying the representative
vortex-like input state in Eq.~\eqref{eq:lg_input} and the short-distance factorization
in Eq.~\eqref{eq:split_step_factorization}. The purpose of this appendix is not to claim
that Eq.~\eqref{eq:lg_input} is an exact propagated solution of the finite-width beam
equation, but rather to make explicit why it is a natural and well-defined initial condition,
and why Eq.~\eqref{eq:split_step_factorization} is a controlled short-distance approximation.

\subsection{Vortex-like input state}
\label{app:vortex_input}

The effective beam equation,
\begin{equation}
i\partial_z\Psi=
\left[
-\frac{1}{2k_0}\nabla_\perp^2\,\mathbb{I}
+\Omega\rho\,\sigma_3
\right]\Psi,
\label{eq:app_beam_eq}
\end{equation}
is intended to propagate a chosen finite-width input spinor. In this setting, the Gaussian input
in Eq.~\eqref{eq:gaussian_input} and the vortex-like input in Eq.~\eqref{eq:lg_input} are
representative initial states rather than exact stationary eigenmodes of the full propagator.

The vortex-like choice
\begin{equation}
\Psi_{LG}(\rho,\phi,0)=
\mathcal{N}
\left(\frac{\rho}{w_0}\right)^{|m|}
e^{-\rho^2/w_0^2}
e^{im\phi}\chi_0
\label{eq:app_lg_input}
\end{equation}
has the expected properties of a finite-width paraxial beam carrying azimuthal phase winding:

\begin{enumerate}
\item
The factor \(e^{im\phi}\) imposes the topological charge \(m\), so that the field carries the
standard vortex-like angular dependence.

\item
The prefactor \((\rho/w_0)^{|m|}\) guarantees regularity at the symmetry axis. For \(m\neq0\),
the amplitude vanishes as \(\rho^{|m|}\) when \(\rho\to0\), avoiding any singular behavior there.

\item
The Gaussian envelope \(e^{-\rho^2/w_0^2}\) ensures transverse localization and square
integrability.

\item
The polarization spinor \(\chi_0\) remains factored from the spatial part at the input plane,
which is the natural starting point for the subsequent torsion-induced polarization evolution.
\end{enumerate}

If one normalizes the input power to unity and assumes
\(\chi_0^\dagger \chi_0=1\), then
\begin{equation}
\int_0^{2\pi}\!\!\int_0^\infty
\Psi_{LG}^\dagger(\rho,\phi,0)\Psi_{LG}(\rho,\phi,0)\,
\rho\,d\rho\,d\phi=1.
\label{eq:app_norm_condition}
\end{equation}
Substituting Eq.~\eqref{eq:app_lg_input} gives
\begin{equation}
|\mathcal{N}|^2
\,2\pi
\int_0^\infty
\left(\frac{\rho}{w_0}\right)^{2|m|}
e^{-2\rho^2/w_0^2}\,
\rho\,d\rho=1.
\label{eq:app_norm_integral}
\end{equation}
With the change of variable \(u=2\rho^2/w_0^2\), one finds
\begin{equation}
|\mathcal{N}|^2
\frac{\pi w_0^2}{2^{|m|+1}}
\Gamma(|m|+1)=1,
\end{equation}
and therefore
\begin{equation}
\mathcal{N}=
\left[
\frac{2^{|m|+1}}{\pi w_0^2\,\Gamma(|m|+1)}
\right]^{1/2},
\label{eq:app_lg_norm}
\end{equation}
up to an irrelevant global phase.

Thus, Eq.~\eqref{eq:lg_input} is a perfectly regular and normalizable vortex-like input state.
Its role in the theory is that of a physically transparent initial condition from which the torsion-induced
finite-width evolution may be generated.

\subsection{Short-distance factorization of the propagator}
\label{app:short_distance_factorization}

Starting from Eq.~\eqref{eq:app_beam_eq}, define the effective Hamiltonian
\begin{equation}
H=
-\frac{1}{2k_0}\nabla_\perp^2\,\mathbb{I}
+\Omega\rho\,\sigma_3.
\label{eq:app_hamiltonian}
\end{equation}
The exact formal solution is
\begin{equation}
\Psi(z)=e^{-izH}\Psi(0)
=
\exp\!\left[
z\left(
i\frac{1}{2k_0}\nabla_\perp^2
-i\Omega\rho\,\sigma_3
\right)
\right]\Psi(0).
\label{eq:app_exact_propagator}
\end{equation}
Now introduce the operators
\begin{equation}
A=i\frac{z}{2k_0}\nabla_\perp^2,
\qquad
B=-iz\Omega\rho\,\sigma_3.
\label{eq:app_AB}
\end{equation}
Then Eq.~\eqref{eq:app_exact_propagator} is simply
\begin{equation}
\Psi(z)=e^{A+B}\Psi(0).
\end{equation}

Since both \(A\) and \(B\) are of order \(z\), the Baker--Campbell--Hausdorff expansion gives
\begin{equation}
e^A e^B
=
\exp\!\left(
A+B+\frac12[A,B]+\cdots
\right),
\label{eq:app_bch}
\end{equation}
so that
\begin{equation}
e^{A+B}=e^A e^B+\mathcal{O}(z^2),
\label{eq:app_factorization_order}
\end{equation}
provided the commutator term remains subleading over the propagation interval considered.

The key point is that the diffraction operator and the torsional operator do not commute.
Indeed, in cylindrical coordinates,
\begin{equation}
\nabla_\perp^2=
\partial_\rho^2+\frac{1}{\rho}\partial_\rho+\frac{1}{\rho^2}\partial_\phi^2.
\label{eq:app_laplacian}
\end{equation}
Acting on a test function \(f(\rho,\phi)\), one finds
\begin{align}
[\nabla_\perp^2,\rho]f
&=
\nabla_\perp^2(\rho f)-\rho\nabla_\perp^2 f
\nonumber\\
&=
\left(
2\partial_\rho+\frac{1}{\rho}
\right)f.
\label{eq:app_comm_lap_rho}
\end{align}
Therefore,
\begin{equation}
[A,B]
=
\frac{\Omega z^2}{2k_0}
\left(
2\partial_\rho+\frac{1}{\rho}
\right)\sigma_3,
\label{eq:app_comm_AB}
\end{equation}
which is nonzero. This shows that the factorization is not exact, but rather a controlled
short-distance approximation.

To first order in \(z\), however, one may write
\begin{equation}
\Psi(z)\approx
e^{i\frac{z}{2k_0}\nabla_\perp^2}
e^{-iz\Omega\rho\sigma_3}
\Psi(0),
\label{eq:app_split_step}
\end{equation}
which is precisely Eq.~\eqref{eq:split_step_factorization}. The first exponential describes
ordinary paraxial diffraction, while the second contains the torsion-induced relative phase
between the two circular components.

If diffraction is neglected over a sufficiently short propagation interval, Eq.~\eqref{eq:app_split_step}
reduces further to the local-phase evolution
\begin{equation}
\Psi(\rho,\phi,z)\approx
e^{-iz\Omega\rho\sigma_3}\Psi(\rho,\phi,0).
\label{eq:app_local_phase_only}
\end{equation}
For a linearly polarized input,
\begin{equation}
\chi_0=\frac{1}{\sqrt2}
\begin{pmatrix}
1\\
1
\end{pmatrix},
\label{eq:app_linear_spinor}
\end{equation}
and a general vortex-like envelope
\begin{equation}
u_m(\rho,\phi)=
\mathcal{N}
\left(\frac{\rho}{w_0}\right)^{|m|}
e^{-\rho^2/w_0^2}
e^{im\phi},
\label{eq:app_um}
\end{equation}
one obtains
\begin{equation}
\Psi(\rho,\phi,z)\approx
u_m(\rho,\phi)\frac{1}{\sqrt2}
\begin{pmatrix}
e^{-i\Omega\rho z}\\[1mm]
e^{+i\Omega\rho z}
\end{pmatrix}.
\label{eq:app_local_vortex_spinor}
\end{equation}
For \(m=0\), this reduces to the Gaussian short-distance expression used in the main text.
More generally, the azimuthal factor \(e^{im\phi}\) modifies the spatial envelope but does not alter
the local torsion-induced polarization rotation law itself, which remains governed by the same
relative phase \(2\Omega\rho z\).

In this sense, Eq.~\eqref{eq:lg_input} and Eq.~\eqref{eq:split_step_factorization} play distinct but
complementary roles: the first specifies a natural class of finite-width input states, whereas the
second gives the controlled short-distance propagator that isolates the torsional phase imprint on
those inputs.

\section{Consistency checks and limiting cases}
\label{app:checks}

The present formulation inherits the consistency checks of the underlying local model. First, in the torsionless limit $\Omega\to0$, Eq.~\eqref{eq:dispersion_quartic} reduces to the degenerate dispersion $k_z=\omega/c$, Eq.~\eqref{eq:rotation_law} gives $\Delta\theta\to0$, and the beam equation \eqref{eq:spinor_beam} becomes the standard paraxial diffraction equation without helicity splitting. Second, gauge invariance remains manifest because the field strength retains its torsion-independent definition and torsion enters only through covariant derivatives. Third, the eigenvectors of the local polarization matrix are precisely circular combinations, confirming that the torsional coupling is helicity selective.

\section{Order-of-magnitude estimates}
\label{app:estimates}

To connect the minimal dimensionless model with plausible physical scales, it is useful to perform a simple order-of-magnitude translation of the torsion parameter. In defect-rich semiconductor platforms, values in the range $\Omega\sim 10^3$ to $10^4\,\mathrm{m}^{-1}$ may be taken as representative illustrative scales for a weak-torsion regime. For beam radii $\rho\sim 10$ to $20\,\mu\mathrm{m}$, this gives $\Omega\rho\sim 10^{-2}$ or below, which is consistent with the local expansion underlying the optical analysis \cite{BelichSilva2026,BakkeMoraes2012,BakkeMoraes2014}.

In the same parameter window, the effective birefringence
\begin{equation}
\Delta n=\frac{2c\Omega\rho}{\omega}
\end{equation}
is expected to be very small, typically in the range $10^{-9}$ to $10^{-8}$ for visible or near-infrared frequencies. The corresponding polarization rotations over millimeter-scale propagation lengths are then also small, potentially reaching the millidegree range under favorable conditions. These numbers should be interpreted with caution: they are not intended as a quantitative prediction for a specific material platform, but only as indicative scales showing that the weak-torsion regime adopted in the theory is internally consistent and that the resulting signals are not manifestly unreasonable from an experimental point of view.

For this reason, the dimensionless values of $\Gamma_0$ used in the figures should be read mainly as illustrative choices that make the internal consequences of the minimal torsion model visually transparent. They are not intended as a one-to-one fit to a given experimental system. A genuine quantitative comparison would require a platform-specific constitutive model, realistic defect statistics, and a treatment of non-geometric optical effects beyond the present scope.

\section{Regularity of the spiral geometry near the symmetry axis}
\label{app:axis_regularity}

A possible concern about the metric employed in the present work is its behavior near the symmetry axis \(\rho=0\), where cylindrical coordinates become degenerate. It is therefore useful to make explicit that the apparent subtlety is only coordinate-based and does not correspond to a genuine geometric singularity of the spiral background.

The first important point is interpretational. The line element
\begin{equation}
dl^2=d\rho^2+\rho^2 d\phi^2+\left(dz+\Omega\rho^2 d\phi\right)^2
\label{eq:app_spiral_metric}
\end{equation}
should be understood as the effective geometry of a \emph{continuous homogeneous distribution of parallel screw dislocations}, rather than as the geometry of a single isolated screw defect. This distinction is essential near the axis. In the geometric theory of defects, the metric of an isolated screw dislocation carries a torsion flux concentrated on the defect line. By contrast, the ``spiral'' metric in Eq.~\eqref{eq:app_spiral_metric} is introduced precisely as the coarse-grained geometry associated with a continuous distribution, for which the torsional content is spread throughout the medium rather than concentrated at \(\rho=0\).

The second point is that the metric itself is regular on the axis. Using Cartesian coordinates,
\begin{equation}
x=\rho\cos\phi,\qquad y=\rho\sin\phi,
\end{equation}
one has the standard identity
\begin{equation}
\rho^2 d\phi = x\,dy-y\,dx.
\label{eq:app_cartesian_identity}
\end{equation}
Substituting Eq.~\eqref{eq:app_cartesian_identity} into Eq.~\eqref{eq:app_spiral_metric}, the line
element becomes
\begin{equation}
dl^2 = dx^2+dy^2+\bigl(dz+\Omega(x\,dy-y\,dx)\bigr)^2.
\label{eq:app_cartesian_metric}
\end{equation}
In this form, the geometry is manifestly smooth at \((x,y)=(0,0)\), since the one-form
\(x\,dy-y\,dx\) is regular on the full plane. Therefore, the apparent difficulty at \(\rho=0\) is
simply the usual coordinate degeneracy of cylindrical coordinates, not a physical singularity of
the metric.

The same conclusion follows from the orthonormal coframe
\begin{equation}
\theta^1=d\rho,\qquad
\theta^2=\rho\,d\phi,\qquad
\theta^3=dz+\Omega\rho^2 d\phi,
\label{eq:app_coframe}
\end{equation}
for which
\begin{equation}
dl^2=(\theta^1)^2+(\theta^2)^2+(\theta^3)^2.
\end{equation}
The torsion two-form associated with the third coframe component is
\begin{equation}
T^3=d\theta^3
=2\Omega\,\rho\,d\rho\wedge d\phi
=2\Omega\,\theta^1\wedge\theta^2.
\label{eq:app_torsion_twoform}
\end{equation}
Equation~\eqref{eq:app_torsion_twoform} shows that the effective torsion is finite and uniform in
the orthonormal frame, rather than singular on the axis. In particular, no divergence appears as
\(\rho\to0\).

This regularity is also reflected in the theory's optical sector. The local helicity splitting derived in the main text,
\begin{equation}
\Delta k_z=2\Omega\rho,
\label{eq:app_delta_k}
\end{equation}
vanishes smoothly at the axis. The associated local polarization rotation law,
\begin{equation}
\Delta\theta(\rho,L)=\Omega\rho L,
\label{eq:app_delta_theta}
\end{equation}
also tends continuously to zero as \(\rho\to0\). Likewise, in the effective paraxial beam model, the torsion-induced coupling term is proportional to
\begin{equation}
\Omega\rho\,\sigma_3,
\end{equation}
and therefore disappears smoothly on the symmetry axis. Thus, neither the geometric background nor the beam-level optical response develops any singular behavior at \(\rho=0\) within the present continuum description.

It is nevertheless important to keep the model's physical meaning explicit. The spiral metric
is a coarse-grained effective geometry, valid above the microscopic scale set by the defect core
size and/or the characteristic spacing between dislocations. The continuum theory is therefore not
intended to resolve atomistic core physics in an infinitesimal neighborhood of the axis. This
limitation does not affect the observables studied in the main text, since the beam-level
quantities are controlled by transverse scales of order \(w_0\), which are much larger than the
microscopic cutoff of the defect description.

In this sense, the behavior near \(\rho=0\) is fully controlled within the effective model. The
symmetry axis is not a geometric singularity of the spiral background, but simply the axis of a
regular coarse-grained torsional geometry. The optical response is equally regular there, with the
torsion-induced splitting and polarization rotation vanishing smoothly on the axis.

\end{document}